\documentclass{ieeeaccess}
\usepackage{graphicx}
\usepackage[font={sf,scriptsize},labelfont={bf,color=accessblue},caption=false]{subfig} 
\usepackage{amsmath}
\usepackage{cite}
\usepackage{colortbl}
\usepackage{acronym}
\usepackage{xcolor}
\usepackage{bm}
\usepackage{balance}
\usepackage{amssymb}
\usepackage{amsthm}
\usepackage{diagbox}
\usepackage{multirow}
\usepackage{booktabs}
\usepackage{float}
\graphicspath{ {./images/} }
\usepackage[utf8]{inputenc}
\usepackage[T1]{fontenc}
\usepackage{color}
\usepackage{amsfonts}
\usepackage{algorithmic}
\usepackage{textcomp}

\begin{document}
	\bstctlcite{IEEEexample:BSTcontrol}
\history{Date of publication xxxx 00, 0000, date of current version xxxx 00, 0000.}
\doi{xxxx}

\title{
	%A Channel Estimation Approach for Hybrid Massive MIMO Systems based on Geospatial Data and Machine Learning
	A Deep Learning and Geospatial Data based Channel Estimation Technique for Hybrid Massive MIMO Systems
}
\author{
%	\uppercase{First A. Author}\authorrefmark{1}, \IEEEmembership{Fellow, IEEE},	\uppercase{Second B. Author\authorrefmark{2}, and Third C. Author, Jr}.\authorrefmark{3}, \IEEEmembership{Member, IEEE}
	Xiaoyi Zhu \IEEEmembership{Student Member, IEEE}, 
	Asil Koc \IEEEmembership{Student Member, IEEE}, 
	Robert Morawski, 
	Tho Le-Ngoc \IEEEmembership{Life Fellow, IEEE}, 
}
\address{Department of Electrical and Computer Engineering, McGill University, Montreal, QC H3A 0G4, Canada}
\tfootnote{This work was supported in part by Huawei Technologies Canada and in part by the Natural Sciences and Engineering Research Council of Canada (NSERC). This article is presented in part in IEEE International Conference on Communications (ICC 2021), Montreal, Canada,
June 2021 in\cite{ASIL_XIAOYI_ICC}.}

\markboth
{X. Zhu \headeretal: Deep Learning and Geospatial Data based Channel Estimation for Massive MIMO}
{X. Zhu \headeretal: Deep Learning and Geospatial Data based Channel Estimation for Massive MIMO}

\corresp{Corresponding author: Asil Koc (e-mail: asil.koc@mail.mcgill.ca).}

\begin{abstract}
This paper presents a novel channel estimation technique for the multi-user massive multiple-input multiple-output (MU-mMIMO) systems using angular-based hybrid precoding (AB-HP). The proposed channel estimation technique generates group-wise channel state information (CSI) of user terminal (UT) zones in the service area by deep neural networks (DNN) and fuzzy c-Means (FCM) clustering. The slow time-varying CSI between the base station (BS) and feasible UT locations in the service area is calculated from the geospatial data by offline ray tracing and a DNN-based path estimation model associated with the 1-dimensional convolutional neural network (1D-CNN) and regression tree ensembles. Then, the UT-level CSI of all feasible locations is grouped into clusters by a proposed FCM clustering. Finally, the service area is divided into a number of non-overlapping UT zones. Each UT zone is characterized by a corresponding set of clusters named as UT-group CSI, which is utilized in the analog RF beamformer design of AB-HP to reduce the required large online CSI overhead in the MU-mMIMO systems. Then, the reduced-size online CSI is employed in the baseband (BB) precoder of AB-HP. 
Simulations are conducted in the indoor scenario at 28~GHz and tested in an AB-HP MU-mMIMO system with a uniform rectangular array (URA) having  $16 \times 16=256$ antennas and $22$ RF chains. 
{Illustrative results indicate that $91.4\%$ online CSI can be reduced by using the proposed offline channel estimation technique as compared to the conventional online channel sounding. The proposed DNN-based path estimation technique produces same amount of UT-level CSI with runtime reduced by $65.8\%$ as compared to the computationally expensive ray tracing. The imperfection of UT-level CSI introduced by the DNN-based path estimation technique is mitigated by the FCM clustering technique, where the AB-HP using offline UT-group CSI generated by the DNN-based channel estimation model and ray tracing based model for the RF beamformer achieves, respectively, $98.7\%$ and $99.1\%$ sum-rate performance of the fully digital precoding (FDP) technique using full-size online CSI.}

\end{abstract}


\titlepgskip=-15pt

\maketitle

\section{Introduction}
The next generation communication networks experience the severe challenge of an increasing number of connections between the base station (BS) and the dense-deployed user terminals (UTs). It brings a huge demand for high-speed data transmission despite a severe shortage in bandwidth resources. 
{Massive multi-input multi-output (mMIMO) systems with an extremely large number of antenna elements can significantly improve the spectral and energy efficiency in millimeter-wave (mmWave) frequencies via focusing narrow and high-gain beams towards a small region and thus increase the overall data rate and channel capacity via spatial multiplexing\cite{larsson2014massive}. 
For downlink transmission, multi-user mMIMO (MU-mMIMO) using hybrid precoding (HP) is prevalently deployed at the BS to tackle above-mentioned challenges, where HP supports a large number of antenna elements via a concatenation of a radio frequency (RF) beamformer, few power-hungry RF chains and a low-dimension baseband precoder\cite{Mass_MIMO_Hybrid_Survey,Mass_MIMO_Hyb_Survey,Mass_MIMO_Hybrid_Survey_2}.

Configuring the hybrid precoder requires the complete knowledge of channel state information (CSI) for both RF beamformer and baseband precoder regarding a massive number of antenna elements. 
Considering the configuration parameters of the RF beamformer design, there are two possible strategies: i) using fast time-varying instantaneous CSI obtained by conventional online channel sounding\cite{7982782,jiang2016joint,maleki2018hybrid}, or ii) using slow time-varying CSI such as the angular information and the channel covariance matrix \cite{ASILAccess,ASIL_URA_VTC,ASIL_MC_PIMRC,6542746,ASIL_Subconnected_GC,ASIL_PSO_PA_WCNC}. In the first approach where the full-size fast time-varying instantaneous CSI is utilized as input parameters for the hybrid precoder design, a large-size CSI training and feedback overhead is required during online connections. However, in the second approach where the RF beamformer is designed based on the offline slow time-varying CSI, only the small-dimensional fast time-varying instantaneous CSI for the baseband precoder is required for the hybrid precoder configuration.}

\subsection{Related works}
{Channel estimation is essential for MU-mMIMO systems to perform the interference management and to ensure the reliable transmission with linear precoding.}
{In \cite{7982782}, a novel channel estimation approach for the hybrid precoding in MU-mMIMO systems is proposed to estimate the angle information and gain information independently by decomposing the instantaneous CSI, where the RF beamformer is constructed based on the angle-of-arrival (AoA) information. Joint user scheduling and MU-MIMO HP algorithm for mmWave mMIMO systems are discussed in \cite{jiang2016joint}, where the RF beamformer in the BS adopts the channel matrix derived from instantaneous CSI via singular value decomposition (SVD). In \cite{maleki2018hybrid}, the size of the feedback CSI overhead for the RF beamformer in a hybrid processing scheme is practically reduced by quantising the phase entries with limited available bits while achieving satisfactory data rate performance. However, all above studies cannot completely remove the pilot and feedback overhead for the RF beamformer because they use the full-size instantenous CSI. 

In terms of the second scheme with slow time-varying CSI, \cite{6542746} presents a joint spatial division multiplexing (JSDM) technique that serves user locations that have similar channel covariance matrices at a group level. As the channel covariance matrix for each group changes extremely slowly compared to the coherence time of the instantaneous channel matrix, the required online CSI for the RF beamformer is notably decreased by exploiting the groupwise channel covariance information obtained via low protocol overhead. However, this scheme still relies on online channel sounding for the groupwise channel covariance matrix. In \cite{ASILAccess,ASIL_URA_VTC,ASIL_MC_PIMRC}, an angular-based HP (AB-HP) technique is proposed for mMIMO systems based on user groups, where the input parameters of the RF beamformer are the groupwise angle-of-departure (AoD) information. However, this scheme simply assumes that the AoD information for each user group are known.}

{Therefore, obtaining the slow time-varying CSI between the BS and UT groups without any knowledge of the online channel, especially estimating the angle information, is of great importance to fully drop the large-size CSI overhead required for the RF beamformer. In \cite{gustafson2011directional}, the CSI between the BS and the UT (i.e., the directional properties of the channel) is estimated by firstly conducting real-life channel measurements in the environment for the original channel matrix and then extracting the angle information from it via space alternating generalized expectation (SAGE) maximization. The CSI has a high accuracy comparing to CSI obtained by online channel sounding, but the channel measurement campaign is extremely inflexible and time-consuming if the environment and antennas have any change.  In \cite{fugen2005characterization}, a channel estimation framework containing the deterministic ray tracing and the cluster extraction is presented. The deterministic ray tracing overcomes the inflexibility in changeable environment and does not need any knowledge of antennas, but it could be computationally complex if multiple propagation mechanisms are considered and could not achieve a accurate CSI if propagation mechanisms are not correctly assumed.} Deep learning has been introduced in UT-level CSI acquisition because of its strong power to predict unknown variables from measured data. \cite{navabi2018predicting} leverages a feedforward neural network (FFNN) to predict the range of AoA and AoD for the dominant paths of each UT based on data directly collected at the BS. In \cite{8485631}, a deep neural network (DNN) based framework composed of a series of parallel multilayer classifiers is proposed for high-precision AoA/AoD estimation. However, none of them explores the relationship between the environment and the UT-level CSI. As the UT-level CSI depends on the layout of the environment, it is reasonable to generate the UT-level CSI using a data-driven artificial intelligence (AI) model learning from a large amount of three-dimensional (3D) geospatial data of the environment and the related propagation rules.
Meanwhile, the COST 2100 stochastic channel model in\cite{liu2012cost} produces the clusterwise CSI in terms of UT-level CSI for densely deployed UTs within the scope of the entire area by clustering. Nonetheless, conventional clustering techniques such as the KPowerMeans\cite{li2017cluster}, the k-means\cite{moayyed2019clustering} and the k-nearest neighbors\cite{zhang2016interdisciplinary}, are not robust enough to imperfect UT-level CSI produced by the deep learning based technique. Fuzzy clustering such as UK-means in\cite{chau2006uncertain} and the conventional fuzzy c-means (FCM) clustering algorithm in\cite{bezdek2013pattern}, can be promising solution to combat the weakness of the imperfect UT-level CSI.

\subsection{Contributions and organization}
In this paper, a learning-based channel estimation approach for the MU-mMIMO hybrid precoding RF beamformer has been presented. The preliminary UT-level CSI of a selected set of UTs is generated by offline ray tracing based on known geospatial data and then used for a DNN learning the correlation between the UT-level CSI and the geospatial data. The UT-level CSI produced by offline ray tracing and the DNN-based path estimation technique forms cluster-level CSI by a proposed FCM clustering algorithm. The proposed UT grouping algorithm further groups feasible UT locations (in a considered service area) into non-overlapping UT zones and calculates the corresponding UT-group CSI based on cluster-level CSI. Finally, the RF beamformer in the MU-mMIMO system is configured using offline UT-group CSI instead of online channel sounding to avoid large CSI overhead. 

{The main contributions are summarized below:
	\vspace{-0.5ex}
\begin{itemize} 
\item \textbf{Deep learning model for UT-level CSI acquisition:} The ray tracing in the preliminary work is computationally expensive even on the assumption that only two propagation mechanisms are considered when calculating the UT-level CSI for each feasible UT location. This paper introduces a deep learning model for the UT-level CSI acquisition, which generates same amount of slow time-varying UT-level CSI for feasible UT locations in the service area with runtime significantly reduced. We provide extra analysis and conduct extensive simulations for the comparison between the deep-learning approach and the non-deep-learning approach.
\item \textbf{FCM Clustering for imperfect UT-level CSI:} Distinct from hard clustering techniques used for stochastic channel models in \cite{liu2012cost,li2017cluster,moayyed2019clustering,zhang2016interdisciplinary}, the proposed FCM clustering produces robuster cluster-level CSI by introducing fuzziness between clusters, in order to combat the imperfect UT-level CSI caused by the DNN-based path estimation. Compared to the previous FCM clustering technique presented in the preliminary work, the proposed FCM in this paper further prunes clusters by neglecting UT-level CSI distant from the cluster center when calculating the cluster-level CSI. Therefore, the cluster-level CSI is more representative of UT-level CSI in each cluster.
\item \textbf{UT zone formation:} The offline channel estimation technique provides slow time-varying UT-group CSI of each UT zone in the service area for the RF beamformer, where UT zones are defined as 3D irregular shaped non-overlapping spaces corresponding to the geospatial data of the service area. 
\end {itemize}
}

The rest of this paper is organized as follows. The system model is introduced in Section II. Section III presents the proposed channel modeling method using ray tracing and FCM and its illustrative results in the indoor scenario. In Section IV, we propose two approaches using deep learning for CSI acquisition and the comparison between the ray tracing and the deep learning methods. Section V provides the example of this offline channel modeling for AB-HP design in the MU-mMIMO system. Finally, the paper is concluded in Section VI.

\section{Massive-MIMO hybrid precoding \& geometry-based 3D channel model}
In this section, we present the hybrid precoding architecture and the geometry-based 3D channel model.

\vspace{-1ex}

\subsection{Hybrid Precoding Architecture}
\begin{figure}[t!]
	\includegraphics[width=\columnwidth]{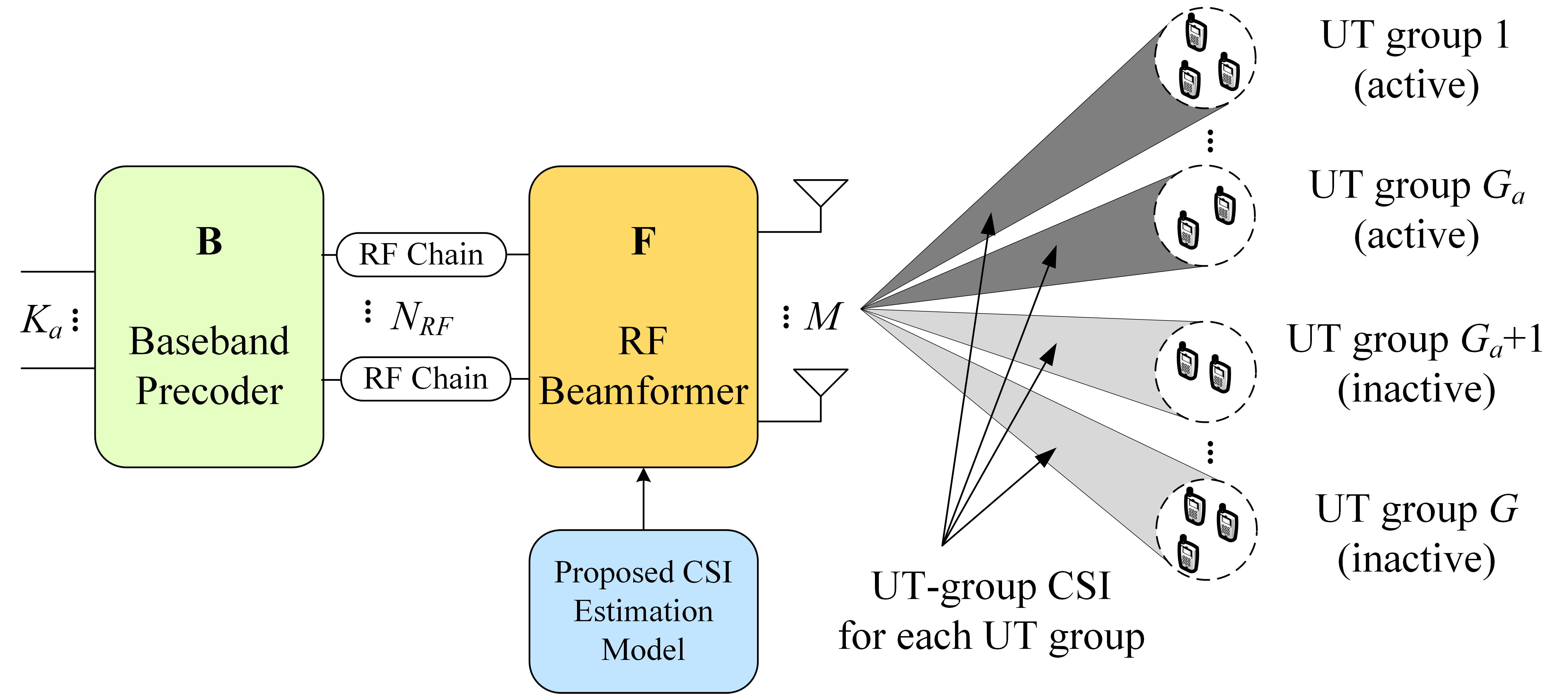}
	\hfill
	\vspace{-3ex}
	\caption{AB-HP technique in the MU-mMIMO systems.}
	\label{fig:systemModel}
	\vspace{-3ex}
\end{figure}

We consider a total of $K$ single-antenna UTs in the service area, where $K_a\leq K$ UTs are active, $K-K_a$ UTs are inactive. The BS is equipped with $M$ antennas deploying the AB-HP technique proposed in\cite{ASILAccess} to serve $K_a$ active UTs as illustrated in Fig.~\ref{fig:systemModel}. 
The downlink signal vector transmitted by the BS is denoted as 
${\bf{s}}={\bf{FBa}}\in \mathbb{C}^{M}$, where 
${\bf{F}}\hspace{-.75ex}\in \hspace{-.75ex}\mathbb{C}^{M\hspace{-.25ex}\times\hspace{-.25ex} N_{RF}}$ is the RF beamformer, 
${\bf{B}}\hspace{-.75ex}\in \hspace{-.75ex}\mathbb{C}^{N_{RF} \hspace{-.25ex}\times \hspace{-.25ex}K_{a}}$ is the BB precoder, 
${\bf{a}}\hspace{-.5ex}\in \hspace{-.5ex}\mathbb{C}^{K_a}\hspace{-.5ex}$ is the downlink data symbol vector to $K_a$ simultaneous UTs, which is encoded by i.i.d Gaussian codebooks (i.e., i.i.d. entries of $\bf a$ follow the distribution of $\mathcal{CN}(0,1)$){\footnote{In the context of this paper, the achievable sum-rate capacity of the hybrid massive MIMO systems have been evaluated in Section V. Hence, we assume that the downlink data symbols are encoded by i.i.d. Gaussian codebooks as in \cite{ASILAccess,ASIL_URA_VTC,ASIL_MC_PIMRC,6542746,ASIL_PSO_PA_WCNC,ASIL_Subconnected_GC,9390210,ASIL_Mobeen_OJCOMS,ASIL_VTC_DAC_ADC}.}}, and $N_{RF}$ is the number of RF chains. 
For the power constraint of $P_T$, we have $\mathbb{E}\left[\| {\bf s}\|_2^2\right]\le P_T$.

Hence, in comparison to the conventional fully digital precoding (FDP), the number of RF chains in the AB-HP technique is reduced from $M$ to $N_{RF}$. 
Meanwhile, in order to support $K_a$ UTs, the number of RF chains should follow $K_a\hspace{-.5ex}\le \hspace{-.5ex}N_{RF}\ll M$. 
Then, the received signal at the $k^{th}$ UT is written as follows:\vspace{-0.5ex}
\begin{equation}
	\begin{aligned}
		r_k &={\bf h}_k^T{\bf F}{\bf {Ba}}+ w_k\\
		&= \underbrace{{\bf h}_k^T{\bf F}{\bf b}_k a_k}_{\text{Desired Signal}} + \underbrace{\sum\nolimits_{q\neq k} {\bf h}_k^T{\bf F}{\bf b}_q a_q }_{\text{Interference}}+ \underbrace{w_k}_{\text{Noise}},
	\end{aligned}
	\vspace{-1ex}
\end{equation}
where 
${\bf h}_k\in\mathbb{C}^{M}$ is the channel vector of the $k^{th}$ UT, 
${\bf b}_k\in\mathbb{C}^{N_{RF}}$ is the $k^{th}$ column of ${\bf B}$, $a_k$ is the $k^{th}$ element of ${\bf a}$, 
and $w_k\hspace{-.25ex}\sim\hspace{-.25ex}\mathcal{CN}\left(0,\sigma^2 \right)$ is the complex Gaussian noise. 
The full channel matrix between the BS and all UTs is defined as ${\bf H}= \left[{\bf h}_1,\cdots, {\bf h}_{K_a}\right] \in\mathbb{C}^{K_a \times M}$. 
The signal-to-interference-plus-noise ratio (SINR) at the $k^{th}$ UT is derived as\cite{ASILAccess}:\vspace{-0.5ex}
\begin{equation}
	\label{eq:SINR}
	\text{SINR}_k = \frac{\left| {\bf h}_k^T{\bf F}{\bf b}_k\right|^2}
	{\left\|{\bf h}_k^T{\bf F}{\bf B}_{\left[k\right]}\right\|^2_2 + \sigma^2},
\end{equation}
where ${\bf B}_{\left[k\right]}\hspace{-0.75ex} = \hspace{-0.75ex}\left[\hspace{-0.25ex}{\bf b}_1,\cdots,{\bf b}_{k-1},{\bf b}_{k+1},\cdots,{\bf b}_{K_a}\hspace{-0.25ex}\right]\in\mathbb{C}^{N_{RF}\times (K_{a}-1)}$.
Afterwards, the average sum-rate of the MU-mMIMO system is obtained by\cite{ASILAccess}:
\begin{equation}
	\label{eq:Rate}
	R_{sum}=\sum_{k=1}^{K_{a}}\mathbb{E}\left[\log_2\left(1+\text{SINR}_k\right)\right] ~ \text{[bps/Hz]}.
\end{equation}
Moreover, we consider a total of $G$ UT groups in the service area, where $K_a$ active UTs are mapped to $G_{a}\leq G$ active UT groups according to the similarity of slow time-varying UT-level CSI and $K_g$ UTs are in the $g^{th}$ UT group such that $K_a=\sum\nolimits_{g=1}^{G_a}K_g$. Then, the AB-HP technique can develop the RF beamformer in blocks ${\bf F} = \left[{\bf F}_1,\cdots{\bf F}_{G_a}\right]$ only using the UT-group CSI, where ${\bf F}_g\in\mathbb{C}^{M\times N_{RF}^{(g)}}$ for each group requires $N_{RF}^{(g)}$ RF chains such that $N_{RF}=\sum\nolimits_{g=1}^{G_a}N_{RF}^{(g)}$. 
	In other words, the AB-HP technique only utilizes the slow time-varying AoD information to develop the RF beamformer $\bf F$.
Afterwards, instead of using the full channel matrix $\bf H$, the BB precoder employs the reduced-size effective channel matrix $\bm{\mathcal{H}}={\bf H}\bf{F}\in\mathbb{C}^{K_a\times N_{RF}}$ seen from the BB-stage. 
Unlike the conventional FDP, the AB-HP technique can significantly reduce the hardware cost/complexity and the CSI overhead. 
In this paper, the development of RF beamformer $\bf F$ and BB precoder $\bf B$ is omitted for focusing on the channel estimation as the main motivation of this paper. Please see \cite{ASILAccess} for the design of RF beamformer $\bf F$ and BB precoder $\bf B$.

\subsection{Geometry-based 3D Channel Model}
According to the geometry-based 3D channel model\cite{ASIL_MC_PIMRC,ASIL_URA_VTC,ASILAccess}, the channel vector between the BS and the $k^{th}$ UT is defined as follows:	
\begin{equation}
	{\bf h}_k =\sum_{j=1}^{N_c}\sum_{l=1}^{L_{j,k}} \sqrt{P_{k,j}} g_{k,j_l}\Lambda\left(\varphi_{k,j_l},\theta_{k,j_l}\right)\bm{\phi}\left(\varphi_{k,j_l},\theta_{k,j_l}\right),
\end{equation}
where $N_c$ is the number of clusters, $L_{j,k}$ is the number of paths in the $j^{th}$ cluster with $j\hspace{-.5ex}=\hspace{-.5ex}1,\cdots,N_c$, $N_L=\sum_{j=1}^{N_c}L_{j,k}$ is the total number of paths, 
$g_{k,j_l}\hspace{-.2ex}\sim\hspace{-.2ex}\mathcal{CN}\left(0,\frac{1}{N_L}\right)$ is the complex path gain of the $l^{th}$ path in the $j^{th}$ cluster, $P_{k,j}$ is the average power of the $j^{th}$ cluster. 
Here, $\Lambda\left(\varphi_{k,j_l},\theta_{k,j_l}\right)$ and, $\bm{\phi}(\varphi_{k,j_l},\theta_{k,j_l})\hspace{-.25ex}\in\hspace{-.25ex}\mathbb{C}^{M}$ denotes the antenna gain and phase response vector, respectively, at the corresponding $(\varphi_{k,j_l},\theta_{k,j_l})$ angle of departure (AoD) direction where the azimuth AoD (AAoD) of the $l^{th}$ path in the $j^{th}$ cluster is $\varphi_{k,j_l}\hspace{-.5ex}\in\hspace{-.5ex}\big[\mu_{k,j}^{\varphi_D}-\sigma_{k,j}^{\varphi_D},\mu_{k,j}^{\varphi_D}+\sigma_{k,j}^{\varphi_D}\big]$, i.e., specified by the mean AAoD $\mu_{k,j}^{\varphi_D}$ and the spread of AAoD $\sigma_{k,j}^{\varphi_D}$. Similarly, the elevation AoD (EAoD) of the $l^{th}$ path in the $j^{th}$ cluster is $\theta_{k,j_l}\hspace{-0.5ex}\in\hspace{-0.5ex}\big[\mu_{k,j}^{\theta_D}-\sigma_{k,j}^{\theta_D},\mu_{k,j}^{\theta_D}+\sigma_{k,j}^{\theta_D}\big]$, i.e., specified by the mean EAoD $\mu_{k,j}^{\theta_D}$ and the spread of EAoD $\sigma_{k,j}^{\theta_D}$.

\section{Channel Estimation based on Geospatial Data and Fuzzy c-Means}
\begin{figure*}[t!]
	\centering
	\includegraphics[width=0.95\linewidth]{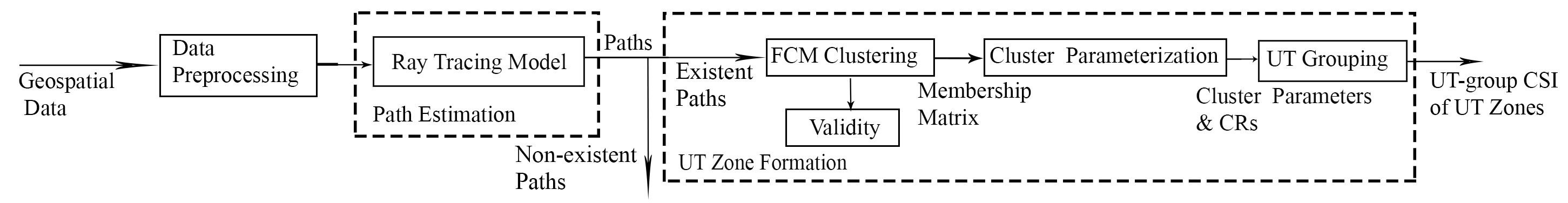}
	\caption{Offline CSI computation based on geospatial data and the FCM technique.}
	\label{fig:Overview}
\end{figure*}

As discussed above, designing the RF beamformer $\bf F$ needs the knowledge of CSI in terms of a large $M$ channel gain and phase response pairs and hence requires a large feedback communications overhead between UTs and the BS if using conventional online channel sounding. To eliminate this requirement, instead of online channel sounding, we propose to perform offline CSI computations based on 3D geospatial data of the service area and FCM technique as shown in Fig.~\ref{fig:Overview}. Firstly, the data preprocessing simplifies the original geospatial data representation to reduce the computational complexity in the subsequent phases. Next, the path estimation generates UT-level CSI in terms of all paths from the known BS position to all feasible UT locations in the service area by offline ray tracing. Finally, the UT zone formation derives all non-overlapping UT zones with their associated CSI via 3 main steps: (i) FCM clustering to form the relationship between existent paths and clusters via the membership matrix, (ii) cluster parameterization to develop the clusters, their regions and parameters, and (iii) UT grouping to establish the non-overlapping UT zones and their associated CSI.

\subsection{Geospatial Data Preprocessing}
{The original geospatial data defines objects by vertices of their planes in Universal Transverse Mercator (UTM) Coodinates System\cite{dma1989}, where curved planes are modeled by several flat planes. For easy data intepretaion and visualization, vertices are re-defined in a relative coordinate system by setting a new origin instead of in the original absolute coordinate system. Then, for computational simplicity, we quantize all vertices by proper resolution to merge objects not affecting the propagation, such as the interior and the exterior walls of buildings.}

\subsection{Path Estimation}
The offline ray tracing generates UT-level CSI for the entire service area via four steps: (i) collect feasible UT locations in the service area from its geospatial data, (ii) produce paths in terms of all objects between the BS and all feasible UT locations, and (iii) compute path parameters according to the trajectory and the path loss model for all paths. For each BS-UT location pair, paths that carry signals arriving at the UT with a power level of 25~dB lower than the strongest signal are labeled as non-existent paths\cite{3gpp2019study} while the remaining existent paths are considered as the existent paths. In this research, only dominant line-of-sight (LoS) propagation, i.e., single-segment paths, and single reflection propagation, i.e., 2-segment paths, are considered in the simulation.

\subsubsection{Feasible UT locations}
{In consideration of LoS propagation and single reflection propagation, feasible UT locations are only selected from UT locations outside of the objects in the service area based on the geospatial data. Besides, in order to reduce the required computational complexity, CSI of feasible UT locations distributed 0.5 meter apart in all dimensions in the service area is adoped for the offline channel estimation.}

\subsubsection{LoS path}
The LoS path has one segment. Its trajectory is obtained by examining line-plane intersections of the segment and other objects. The LoS path is non-existent if the segment contains line-plane intersections on any objects. Otherwise, this LoS path is existent. Its path loss is given by \cite{hoseini2017theoretical}:
\begin{equation}
	\label{eq:friis}
	L_s\hspace{-.25ex}\hspace{-.5ex}=\hspace{-.5ex}-10\hspace{-.25ex}\log_{10}\hspace{-0.75ex}\left[\hspace{-.25ex}\frac{\lambda^2}{A_{m}(4\pi)^2d^\eta}\hspace{-.25ex}\right],
\end{equation}
where $\lambda$ is the wavelength, $\eta$ is the path loss exponent in\cite{viswanathan2018wireless}, $d$ is the length of all segments and $A_{m}$ is the frequency-dependent molecular absorption attenuation given by\cite{hoseini2017theoretical}:
\begin{equation}
	A_{m}=\exp(d\mathrm{k}(f)),
\end{equation}
where $\mathrm{k}(f)$ is the molecular absorption coefficient for frequency $f$ available in\cite{rothman2009hitran}. As the molecular absorption does not affect the propagation loss much at low frequencies, $\mathrm{k}(f)=0$ for low $f$ (e.g., at sub-6~GHz). 

\subsubsection{Reflection path}
The single-reflection path has two segments connecting the BS location, the reflected points, and the UT location. Its trajectory is computed based on the mirror theory. The single-reflection path is non-existent if any of segments passes other objects. Otherwise, this single-reflection path is existent. Its path loss is the sum of (\ref{eq:friis}) and the reflection loss given by\cite{mahafza2017introduction}:
\begin{equation}
	\label{eq:reflection}
	L_r=-20\log_{10}(R_cR_s),
\end{equation}
where $R_c$ is the reflection coefficient from Fresnel’s equations and $R_s$ is the surface roughness factor associated with the material of the interacted objects\cite{yang2013deterministic}.

\subsubsection{Path representation}
{The vector of the $s^{th}$ segment in a path, defined as $(x_s\;y_s\;z_s)$ in Cartesian coordinates, is calculated by subtracting one point from the other in the direction from the BS to the UT. To get the angle information of the $s^{th}$ segment, the segment vector is converted from Cartesian coordinates to spherical coordinates $(\varphi_s\;\theta_s\;d_s)$, given by: 
\begin{equation}
	\begin{bmatrix}
		\varphi_s\\
		\theta_s\\
		d_s
	\end{bmatrix}
	=
	\begin{bmatrix}
		\mathrm{atan2}(y_s/x_s)\\
		\mathrm{atan2}(z_s/\sqrt{x_s^{2}+y_s^{2}})\\
		\sqrt{x_s^2+y_s^2+z_s^2}
	\end{bmatrix},
\end{equation}
where $\mathrm{atan2}$ refers to the four-quadrant inverse tangent\cite{swartzlander2015computer}, $\varphi_s$ is the azimuth angle, $\theta_s$ is the elevation angle, $d_s$ is the segment length. 

Paths are characterized by two parameters, a path vector derived from its segments and the received signal strength (RSS). Using angle and distance information of all segments in the $l^{th}$ path (i.e., $(\varphi_1\;\theta_1\;d_1)$ in the direction of BS to UTs for LoS paths, $(\varphi_1\;\theta_1\;d_1)$ in the direction of the BS to the reflected point and $(\varphi_2\;\theta_2\;d_2)$ in the direction of the reflected point to the UT for single-reflection paths), the $l^{th}$ path is characterized by the vector $\bm{x}_l$, defined as:
\begin{equation}
	\label{eq:MPC}
	\bm{x}_l=[\varphi_{D_l},\theta_{D_l},\varphi_{A_l},\theta_{A_l},\tau_l],
\end{equation}
  where $\varphi_{D_l},\theta_{D_l},\varphi_{A_l},\theta_{A_l}$ are AAoD, EAoD, azimuth angle of arrival (AAoA), elevation angle of arrival (EAoA) ranging in $\left[-\pi,\pi\right]$ rad, respectively, and $\tau_l$ is the delay in second. For the existent LoS path, $\varphi_{D_l}=\varphi_{A_l}=\varphi_1$, $\theta_{D_l}=\theta_{A_l}=\theta_1$, and $\tau_l=d_1/\mathrm{c}$, where $\mathrm{c}$ is the speed of the light. For the existent single-reflection path, $\varphi_{D_l}=\varphi_1$, $\varphi_{A_l}=\varphi_2$, $\theta_{D_l}=\theta_1$, $\theta_{A_l}=\theta_2$, and $\tau_l=(d_1+d_2)/\mathrm{c}$. For the non-existent path, $\varphi_{D_l}=\theta_{D_l}=\varphi_{A_l}=\theta_{A_l}=\tau_l=0$.
  
The RSS of the $l^{th}$ path is $P_l=P_t+G_t+G_r-L$ \cite{5635436}, where $P_t$ is the transmit power, $G_t$ is the transmit antenna gain, $G_r$ is the receive antenna gain, $L$ is the path loss obtained by (\ref{eq:friis}) for the LoS path and by (\ref{eq:friis}) and (\ref{eq:reflection}) for the single-reflection path. For the non-existent path, $P_l=-\infty$.} 

\subsection{UT Zone Formation}
We consider that a total of $N$ paths are present for a total of $K$ feasible UT locations in the considered service area, which are obtained via the path estimation. $N_r$ paths $\mathbf{X}=\bm{[x}_1,\cdots,\bm{x}_{N_r}]$ arrive at UTs (i.e., existent paths) and $N-N_r$ paths are blocked by objects (i.e., non-existent paths). In this module, $N_r$ existent paths are grouped to $N_c$ clusters and $K$ UT locations are divided into $G$ non-overlapping UT zones in the considered service area.

\subsubsection{FCM clustering}
{Using FCM clustering, each existent path can belong to multiple clusters with the membership from $0$ to $1$, where higher membership indicates stronger association with the cluster. The proposed clustering algorithm uses the following iterative process to group $N_r$ existent paths into $N_c$ clusters. At first, it initializes the membership matrix $\mathbf{U}=[u_{lj}] \in \mathbb{R}^{N_c \times N_r}$ where $u_{lj}$ is the membership of the existent path $\bm{x}_l$ in the $j^{th}$ cluster such that $\sum_{j=1}^{N_c}u_{lj}=1$. 
In the $n^{th}$ iteration (where $n=1,2,\cdots$), the $j^{th}$ cluster centroid is calculated based on the RSS, given by:
\begin{equation}
	\label{eq:center}
	\bm{v}_j=\frac{\sum_{l=1}^{N_r}{P_l(u_{lj})^m\bm{x}_l}}{\sum_{l=1}^{N_r}P_l (u_{lj})^m}, 
\end{equation}
where $m>1$ is a fuzzy control constant affecting the cluster fuzziness. Subsequently, elements of the membership matrix $\mathbf{U}$ are updated as\cite{bezdek2013pattern}:
\begin{equation}
	\label{eq:membership}
	u_{lj}=\frac{1}{\sum_{q=1}^{N_c}(\frac{E_{lj}}{E_{lq}})^{\frac{2}{m-1}}}. 
\end{equation}
where $E_{lj}$ is the similarity between the existent path $\bm{x}_l$ and the $j^{th}$ cluster to address the angle periodicity, given by:
\begin{equation}
	E_{lj}\hspace{-0.75ex}=\hspace{-1.5ex}\sqrt{\hspace{-0.5ex}\left|x_{l5}\hspace{-0.5ex}-\hspace{-0.5ex}v_{j5}\right|^2\hspace{-1ex}+\hspace{-0.7ex}\sum_{i=1}^{4}(\mathrm{mod}(\left|x_{li}\hspace{-0.5ex}-\hspace{-0.5ex}v_{ji}\right|\hspace{-0.5ex}+\hspace{-0.5ex}\pi,2\pi)\hspace{-0.5ex}-\hspace{-0.5ex}\pi)^2},\nonumber
\end{equation}
where $x_{li}$ and $v_{ji}$ denote the $i^{th}$ components of vectors $\bm{x}_l$ and $\bm{v}_j$, respectively. Finally, the objective function in the $n^{th}$ iteration is defined as\cite{bezdek2013pattern}:
\begin{equation}
	\label{eq:OF}
	J_n=\sum_{l=1}^{N_r}\sum_{j=1}^{N_c}(u_{lj})^m(E_{lj})^2, \; n\geq 1. 
\end{equation}
The iterative process continues to re-compute (\ref{eq:center})-(\ref{eq:OF}) until convergence, i.e., $\Delta{J}=J_n-J_{n-1}<\epsilon$, where $J_0=\infty$.
}

\subsubsection{Validity}
The number of clusters should be selected to appropriately represent all existent paths without high complexity. Thus, we examine the validity of a number of clusters $c$ to determine the best number of clusters based on the partition coefficient $V_\mathrm{PC}(c)$, the partition entropy $V_\mathrm{PE}(c)$, the partition index $V_\mathrm{SC}(c)$, the separation index $V_\mathrm{S}(c)$, the Xie and Beni's index $V_\mathrm{XB}(c)$, defined as\cite{wu2005cluster}:
\begin{align}
	V_\mathrm{PC}(c)&=\frac{1}{N_r}\sum\nolimits_{j=1}^{c}\sum\nolimits_{l=1}^{N_r}(u_{lj})^2,\\
	V_\mathrm{PE}(c)&=-\frac{1}{N_r}\sum\nolimits_{j=1}^{c}\sum\nolimits_{l=1}^{N_r}u_{lj}\log(u_{lj}),\\
	V_\mathrm{SC}(c)&=\sum_{j=1}^{c}\frac{\sum\nolimits_{l=1}^{N_r}(u_{lj})^m\left\|\bm{x}_l-\bm{v}_j\right\|^2}{L_j\sum\nolimits_{q=1}^{c}\left\|\bm{v}_q-\bm{v}_j\right\|^2},\\
	V_\mathrm{S}(c)&=\frac{\sum\nolimits_{j=1}^{c}\sum\nolimits_{l=1}^{N_r}(u_{lj})^m\left\|\bm{ x}_l-\bm{v}_j\right\|^2}{N_r\min_{j,q}\left\|\bm{v}_q-\bm{v}_j\right\|^2},\\
	V_\mathrm{XB}(c)&=\frac{\sum_{j=1}^{c}\sum_{l=1}^{N_r}(u_{lj})^2\left\|\bm{x}_l-\bm{v}_j\right\|^2}{N_r\min_{j,q}\left\|\bm{v}_q-\bm{v}_j\right\|^2},
\end{align}
where $L_j$ is the number of existent paths in the $j^{th}$ cluster and $\sum_{j=1}^{N_c}L_j=N_r$.
The FCM clustering runs repeatedly for $c \in \left\{2,3,\cdots,N_{C_{max}}\right\}$ to get the cluster prototypes, where the upbound $N_{C_{max}}$ depends on the scenario. A best partition has higher $V_\mathrm{PC}$, and lower $V_\mathrm{PE}$, $V_\mathrm{SC}$, $V_\mathrm{S}$ and $V_\mathrm{XB}$\cite{bezdek2013pattern}.

\subsubsection{Cluster parameterization}
The existent path $\bm{x}_l$ is assigned to the $j^{th}$ cluster if $u_{lj}>u_{lq},q=1,\cdots,j-1,j+1,\cdots, N_c$. Then, the $j^{th}$ cluster contains a path set $\mathbf{X}_j=\left\{\bm{x}_1,\cdots,\bm{x}_{L_j}\right\}$. Instead of using $\mathbf{X}_j$, the $j^{th}$ cluster can be represented by\cite{costa2010multiple}:
\begin{equation}
	\begin{split}
		C_j=[\mu_j^{\varphi_D}, \mu_j^{\theta_D},\mu_j^{\varphi_A},\mu_j^{\theta_A},\mu_j^{\tau}, \mu_j^{P},\sigma_j^{\varphi_D}, \sigma_j^{\theta_D},\\
		\sigma_j^{\varphi_A},\sigma_j^{\theta_A},\sigma_j^{\tau},\sigma_j^{P}],
	\end{split}
\end{equation}
where $\mu_j^{P}$ and $\sigma_j^{P}$ are the mean and the spread of the cluster power calculated by:
\begin{align}
	\label{eq:clusterrss}
	\mu_j^{P}&=\frac{1}{L_j}\sum_{l=1}^{L_j}P_l,\\
	\sigma_j^{P}&=\sqrt{\frac{1}{L_j}\sum_{l=1}^{L_j}(P_l-\mu_j^{P})^2}.
\end{align}
$\mu_j^{a(i)}$ and $\sigma_j^{a(i)}$, where $a=[\varphi_D, \theta_D, \varphi_A, \theta_A, \tau], i=1,\cdots,5$, are the mean and the spread of the angles and the delays, given by:
\begin{align}
	\label{eq:clusterangle}
	\mu_j^{a(i)}&=\frac{\sum_{l=1}^{L_j}x_{li}P_l}{\sum_{l=1}^{L_j}P_l},\\
	\sigma_j^{a(i)}&=\sqrt{\frac{\sum_{l=1}^{L_j}(x_{li}-\mu_j^{a(i)})^2P_l}{\sum_{l=1}^{L_j}P_l}}.
\end{align}
The EAoA and AAoA information in the cluster become redundant for formulating RF beamformer in the BS and can be omitted. 

{We apply path pruning to the original clusters in order to eliminate existent paths (i.e., outliers) that have too large Euclidean distances from their cluster center centroid, as well as to preserve the cluster parameters at the same time\cite{czink2006framework}. For the $j^{th}$ cluster, consider that the original path index set is $\mathcal{C}_{j0}=[\mathcal{I}_1,\cdots,\mathcal{I}_{L_j}]$, where $\mathcal{I}_l$ denotes the index of the $l^{th}$ path. The algorithm first calculates the cluster parameters $C_{j0}$ based on the original paths $\mathcal{C}_{j0}$ and take them as benchmarks. Then, in the $n^{th}$ iteration, the algorithm computes the point-to-center distances for all paths in $\mathcal{C}_{j(n-1)}$, discards the existent path with excessive distance (i.e., the $l^{th}$ existent path if $E_{lj}>E_{mj}$ for $m\neq l\leq L_j$) and creates the new path index set $\mathcal{C}_{jn}=[\mathcal{I}_1,\cdots,\mathcal{I}_{l-1},\mathcal{I}_{l+1},\cdots,\mathcal{I}_{L_j}]$. Afterwards, the new cluster parameters $C_{jn}$ are recalculated by (\ref{eq:clusterrss}) and (\ref{eq:clusterangle}), and compared with the benchmark $C_{j0}$ element by element. If values of the cluster parameters $C_{jn}$ are larger than 95\% of $C_{j0}$ (e.g., $\mu_j^{P_n}>0.95\mu_j^{P_0}$, $\sigma_j^{P_n}>0.95\sigma_j^{P_0}$), the algorithm continues pruning the existent paths in the cluster. Otherwise, the $j^{th}$ cluster takes $C_{j(n-1)}$ as its parameters and the iteration stops. With path pruning, the cluster center centroid moves closer to dense paths in the cluster and the spreads of the clusters are more representative of the majority of paths.
}

\subsubsection{Cluster region (CR) identification}
The $j^{th}$ CR $\mathcal{R}_j$ is the geospatial region that includes all feasible UT locations having existent paths in $\mathbf{X}_j$. Part of $\mathcal{R}_j$ containing feasible UT locations with existent paths having $u_{lj}<0.6$ is defined as fuzzy sub-CR $\mathcal{R}_{F,j}$. The other part containing feasible UT locations with existent paths having $u_{lj}\geq 0.6$ is named the deterministic sub-CR $\mathcal{R}_{D,j}$. Therefore, UTs can receive the signals transmitted via the $j^{th}$ cluster if located in $\mathcal{R}_{D,j}$ while the UT is less accessible if located in $\mathcal{R}_{F,j}$. Note that one cluster region can partially overlap with other cluster regions.

\subsubsection{UT grouping}
Feasible UT locations having the same cluster group $\mathcal{Z}_g=\left\{C_1,\cdots,C_{N_g}\right\}$ are assigned to the $g^{th}$ UT group. 
Feasible UT locations in the $g^{th}$ group further form the geospatial $g^{th}$ UT zone in the service area, which include all CRs of the cluster group $\mathcal{Z}_g$. Note that UT zones do not overlap each other. 
The UT-group CSI of the $g^{th}$ UT zone is represented by the parameters of the cluster group $\mathcal{Z}_g$.

\subsection{Illustrative Results}
\begin{figure}[t!]
	\centering
	\subfloat[Real-life service area.]{
		\includegraphics[width=0.6\linewidth]{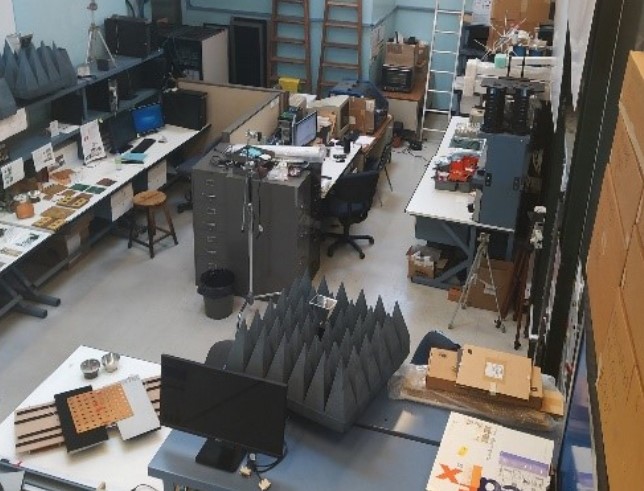}
	}
	\hfill
	\subfloat[3D geospatial model.]{
		\includegraphics[width=0.8\linewidth]{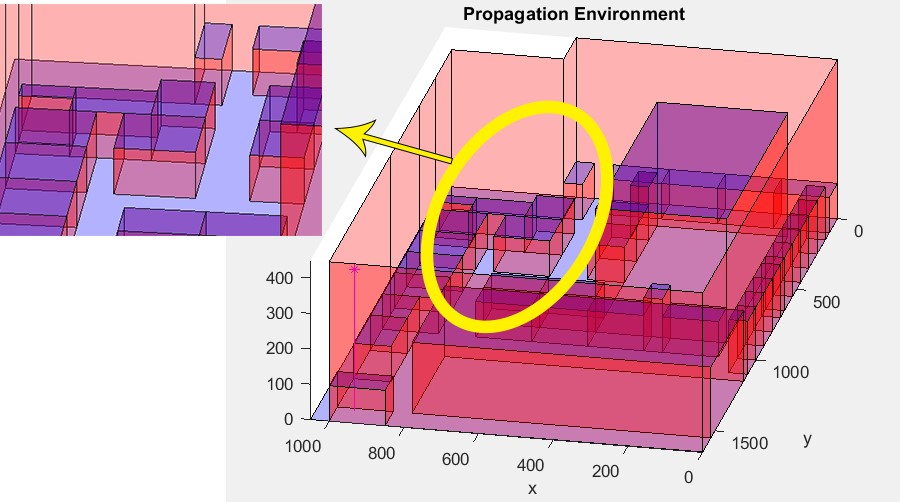}
	}
	\caption{Simulation service area.}
	\label{fig:PE}
\end{figure}

As illustrative results, the proposed channel estimation method is implemented in an indoor scenario at 28~GHz, as shown in Fig.~\ref{fig:PE}, with related parameters and results summarized in Table~\ref{tab:SS} and Table~\ref{tab:indoor results}, respectively. 

\begin{table}[t!]
	\centering
	\caption{Simulation Scenario Parameters.}
	\label{tab:SS}
	\begin{tabular}{|l|c|c|}
	\hline
	\multicolumn{2}{|c|}{Scenario}                                                                                         & Indoor                                              \\ \hline
	\multicolumn{1}{|c|}{\multirow{3}{*}{Layout}} & \begin{tabular}[c]{@{}c@{}}{Volumetric size} \\ (W x L x H)\end{tabular} & 1646 cm x 1036 cm x 450 cm                          \\ \cline{2-3} 
	\multicolumn{1}{|c|}{}                        & Resolution                                                             & 0.5 m                                               \\ \cline{2-3} 
	\multicolumn{1}{|c|}{}                        & Number of Planes                                                       & 205                                                 \\ \hline
	\multirow{2}{*}{BS}                           & Position(x,y,z)                                                        & (950,1570,400) cm                                   \\ \cline{2-3} 
												  & EIRP (dBm)                                                             & 34.5  {(isotropic) } \\ \hline
	UT                                            & Directivity(dBi)                                                       & 2.5 (Omnidirectional)                               \\ \hline
	\multirow{2}{*}{Propagation}                  & Frequency                                                              & 28$\sim$GHz                                         \\ \cline{2-3} 
												  & Path loss exponent                                                     & 2.1                                                 \\ \hline
	\multicolumn{2}{|c|}{Max Number of Clusters}                                                                           & 70                                                  \\ \hline
	\end{tabular}
	\end{table}

\begin{table}[t!]
	\centering
	\caption{Simulation Results}
	\label{tab:indoor results}
	\begin{tabular}{|l|l|} 
	\hline
	Frequency bands                         & 28~GHz                     \\ 
	\hline
	\# of feasible UT locations                               & 3656                      \\ 
	\hline
	\# of Blocked UT locations                      & 458                       \\ 
	\hline
	\# of Linked UT locations                      & 3198                      \\ 
	\hline
	Coverage Ratio                          & 87.47\%                   \\ 
	\hline
	Maximum number of paths in a BS-UT pair & 6                         \\ 
	\hline
	\# of all paths                         & 753136                    \\ 
	\hline
	\# of non-existent paths                & 736504                    \\ 
	\hline
	\# of existent paths                    & 16632                     \\ 
	\hline
	\# of fuzzy paths                       & 349                       \\ 
	\hline
	\# of deterministic paths               & 16283                     \\ 
	\hline
	\# of clusters                          & 56                        \\ 
	\hline
	Cluster Power Range (dBm)               & -58.2629 $\sim$ -21.7646    \\ 
	\hline
	Path Power Range (dBm)                  &-62.2274 $\sim$ -11.9265   \\ 
	\hline
	\# of Region                            & 489                       \\
	\hline
	\end{tabular}
\end{table}

\begin{figure}[t!]
	\centering
	\subfloat[3D view.]{
		\includegraphics[width=0.8\linewidth]{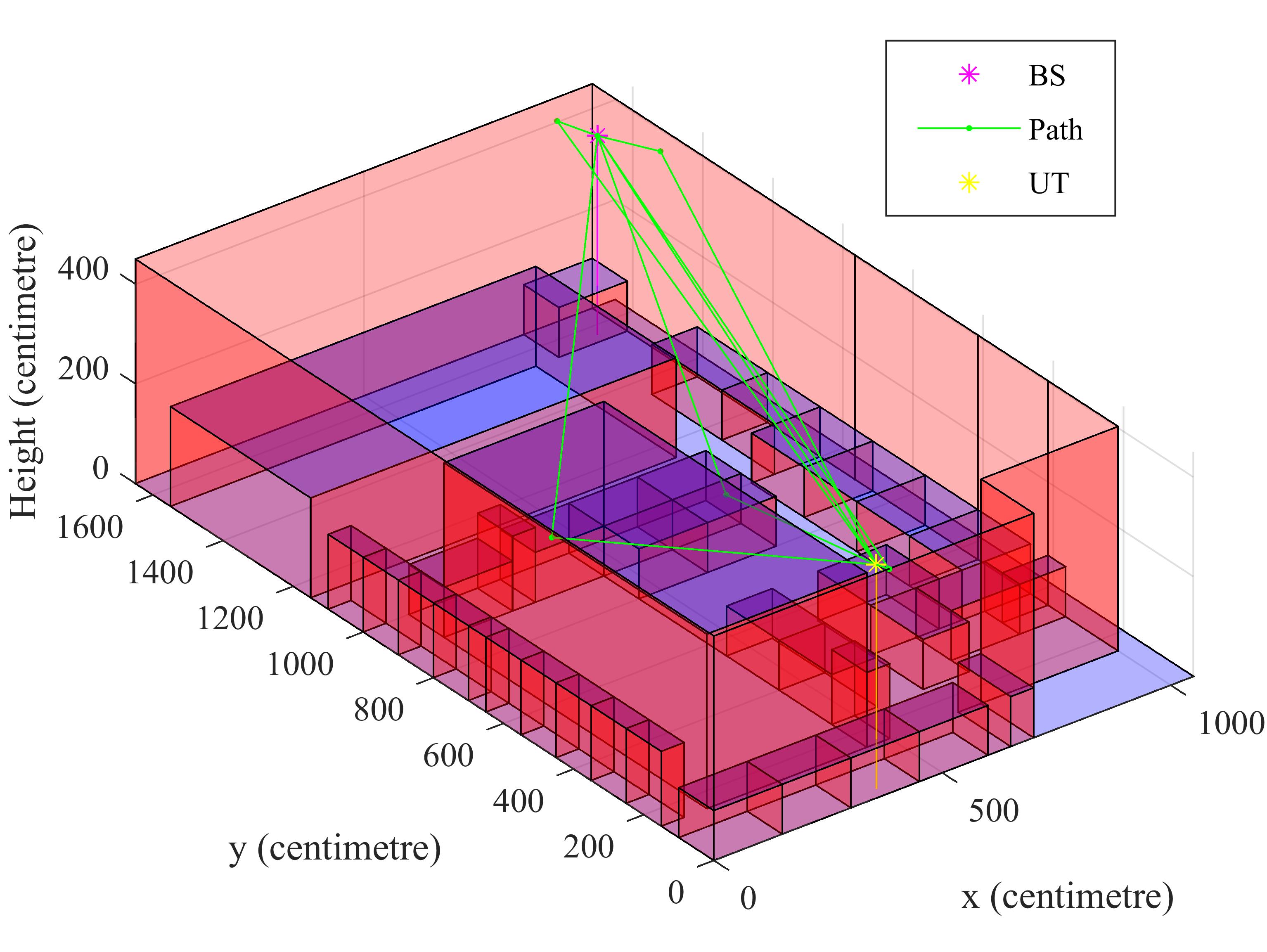}
	}
	\hfill
	\subfloat[Top view.]{
		\includegraphics[width=0.8\linewidth]{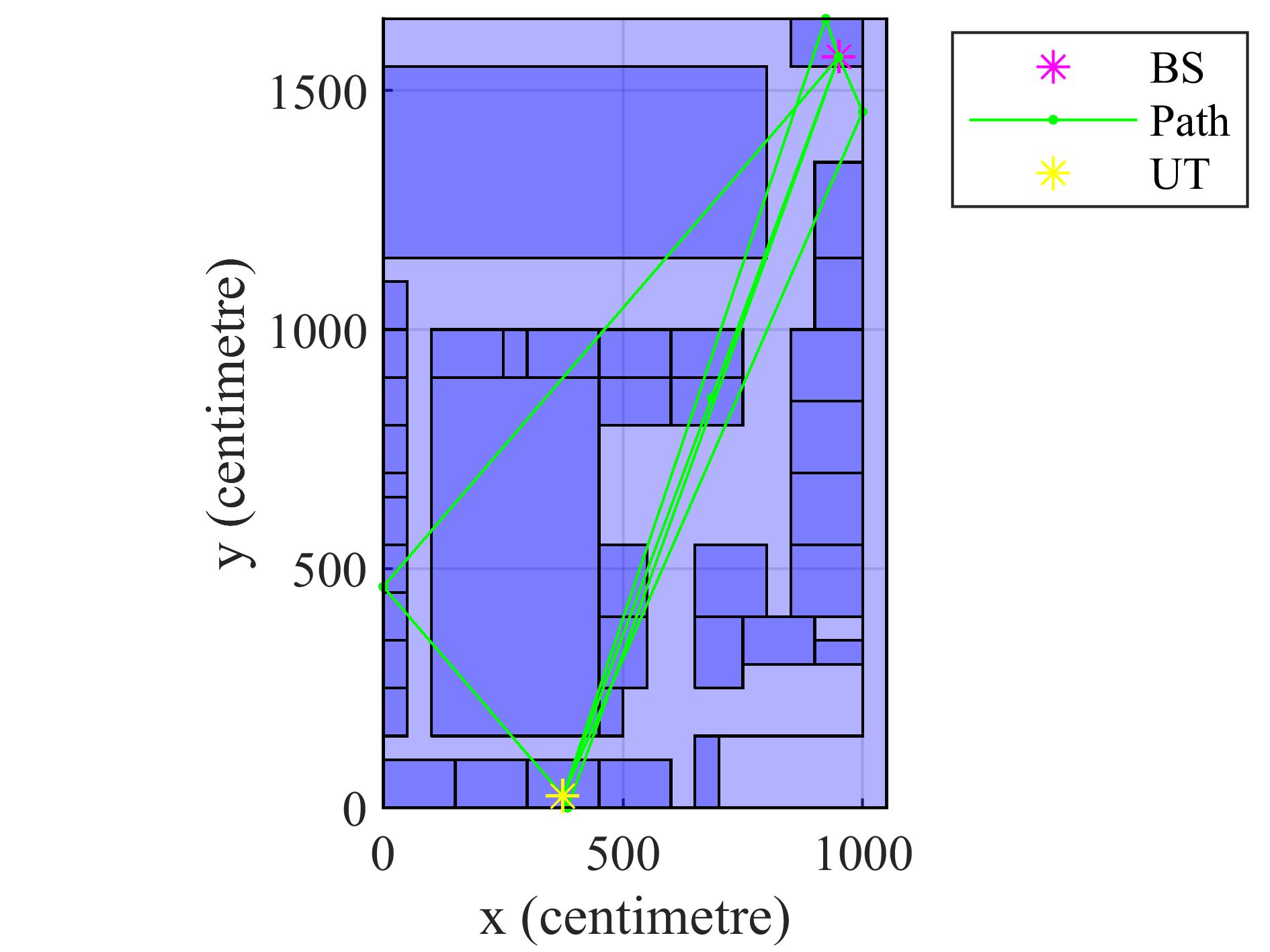}
	}
	\caption{Ray tracing example of a single UT location.}
	\label{fig:RT}
\end{figure}

\begin{figure}[htbp]
	\centering
	\subfloat[Objective function.]{
		\includegraphics[width=\linewidth]{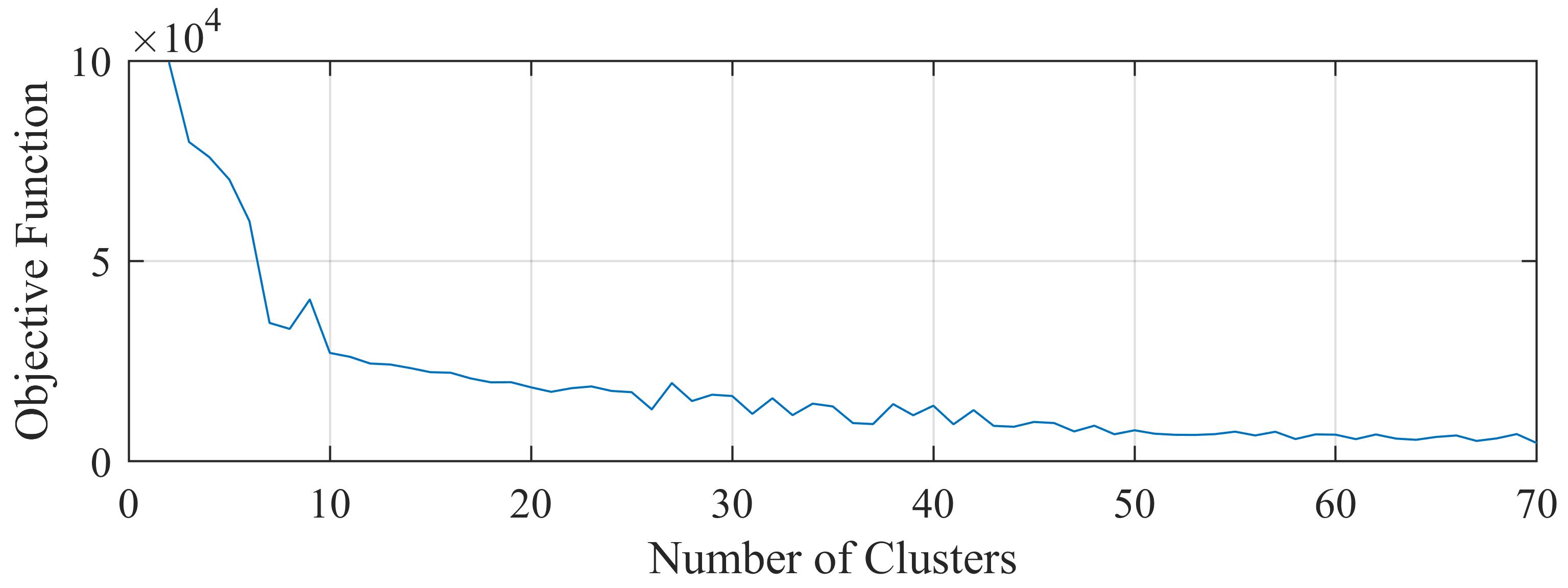}
		\label{fig:28indoorOF}
	}
	\vspace{-0.5ex}
	\hfill
	\subfloat[Partiton coefficient.]{
		\includegraphics[width=\linewidth]{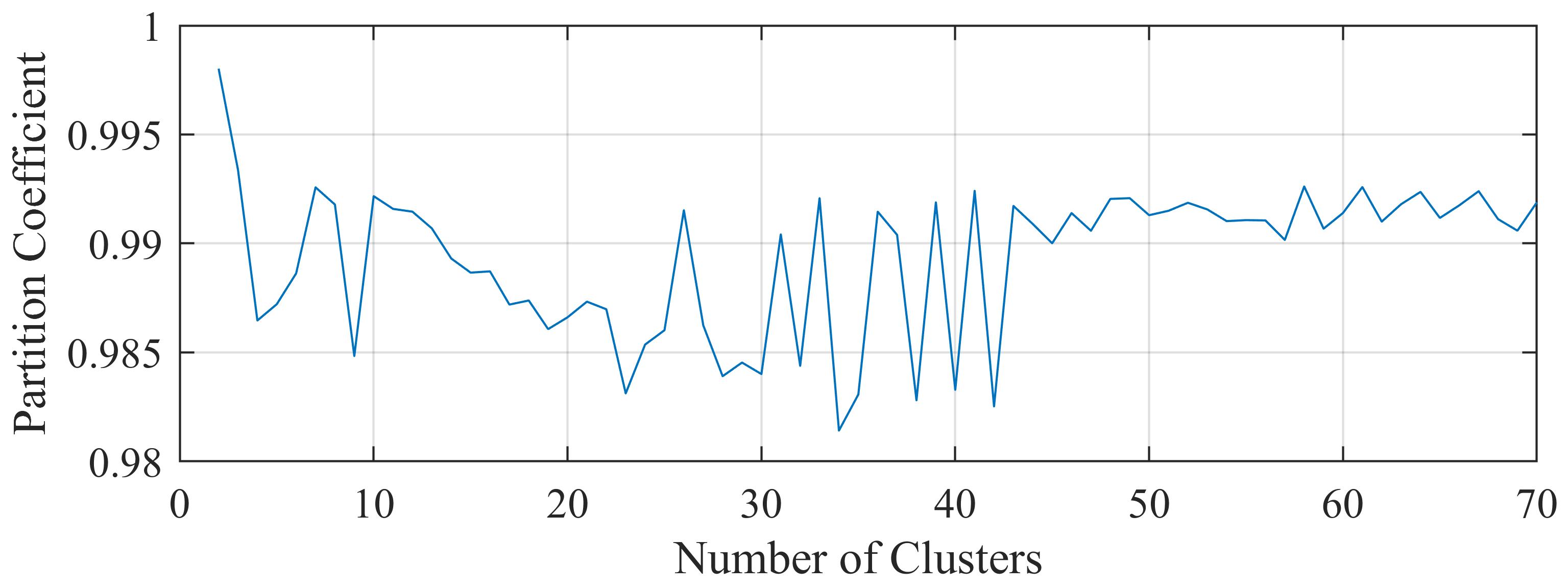}
		\label{fig:28indooorPC}
	}
	\vspace{-0.5ex}
	\hfill
	\subfloat[Partition entropy.]{
		\includegraphics[width=\linewidth]{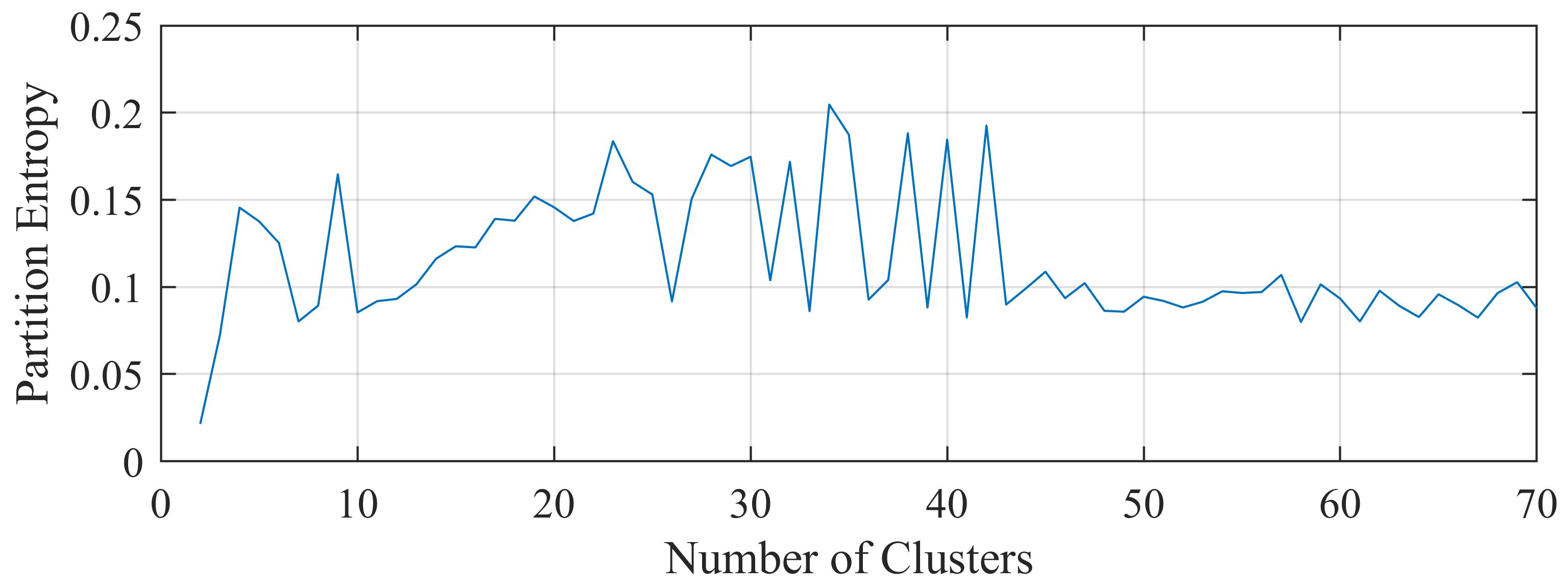}
		\label{fig:28indooorPE}
	}
	\vspace{-0.5ex}
	\hfill
	\subfloat[Partition index.]{
		\includegraphics[width=\linewidth]{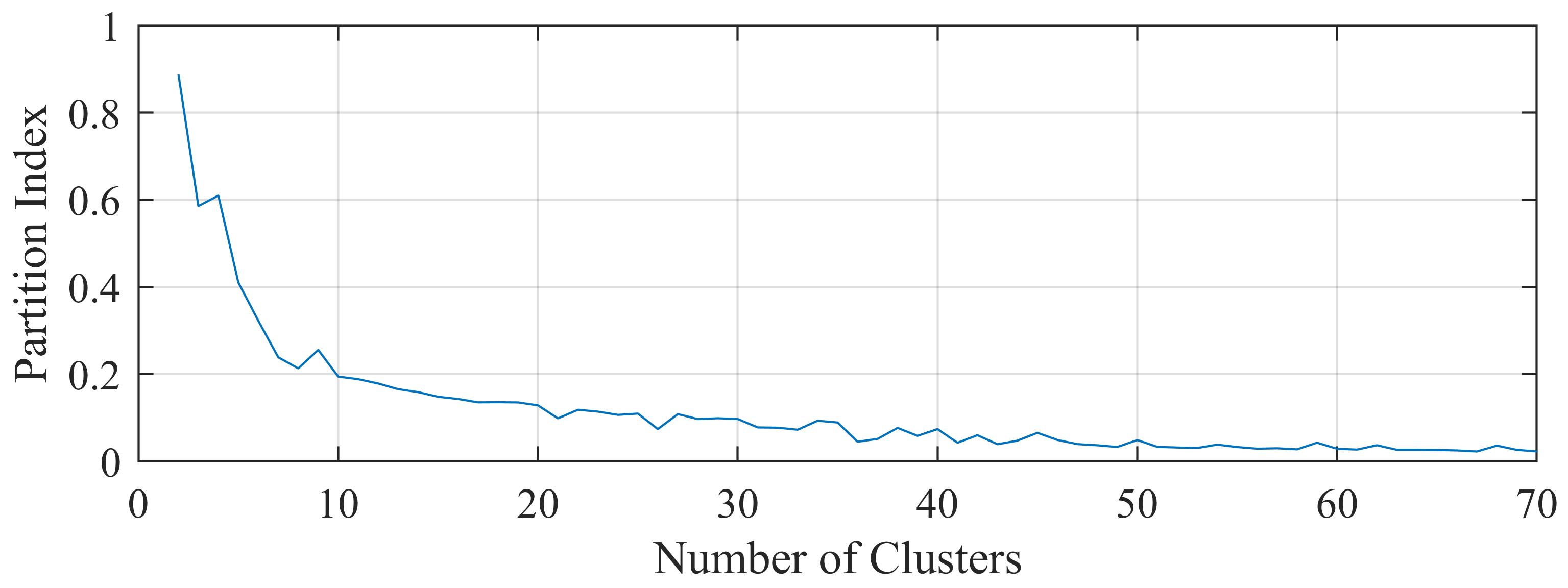}
		\label{fig:28indoorSC}
	}
	\vspace{-0.5ex}
	\hfill
	\subfloat[Separation index.]{
		\includegraphics[width=\linewidth]{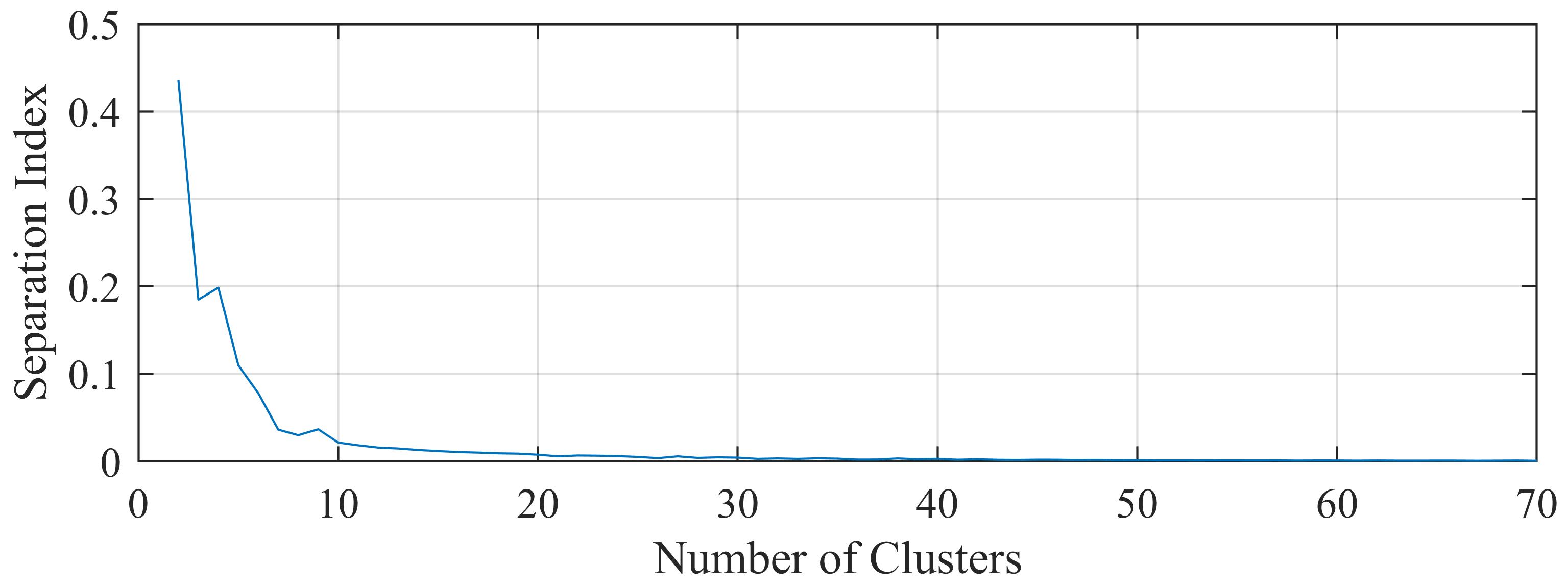}
		\label{fig:28indooorS}
	}
	\vspace{-0.5ex}
	\hfill
	\subfloat[Xie and Beni's index.]{
		\includegraphics[width=\linewidth]{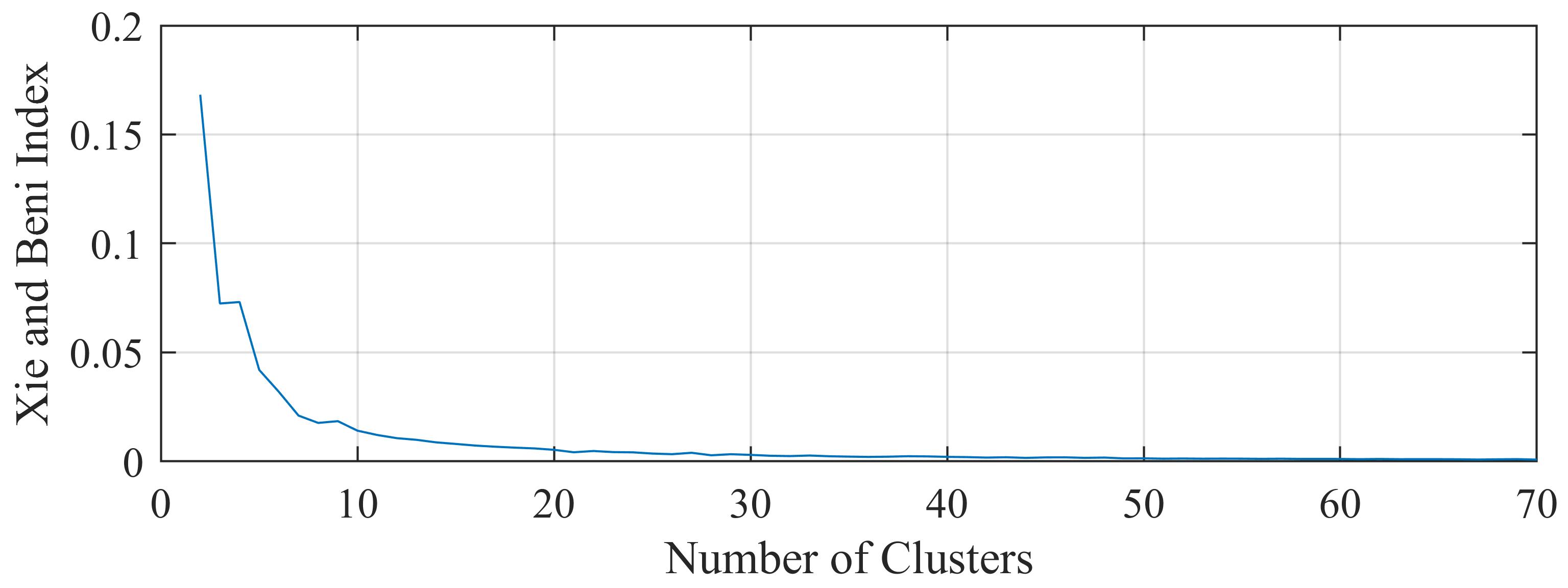}
		\label{fig:28indooorXB}
	}
	\vspace{-0.5ex}
	\hfill
	\vspace{-1ex}
	\caption{Validity indices.}
	\label{fig:VIforIndoor}
\end{figure}

Fig.~\ref{fig:RT} presents the ray tracing example of one UT location, where there are 6 existent paths and 200 non-existent paths.
Fig.~\ref{fig:VIforIndoor} shows the validity results of the indoor 28~GHz scenario. In this case, the best number of clusters is $N_c=58$ because $V_\mathrm{PC}$ and $V_\mathrm{PE}$ reach the local maxima and minima, respectively, while $V_\mathrm{S}$, $V_\mathrm{SC}$ and $V_\mathrm{XB}$ virtually drop to 0.

\begin{figure}[t!]
	\centering
	\subfloat[AoD domain.]{
		\includegraphics[width=0.6\linewidth]{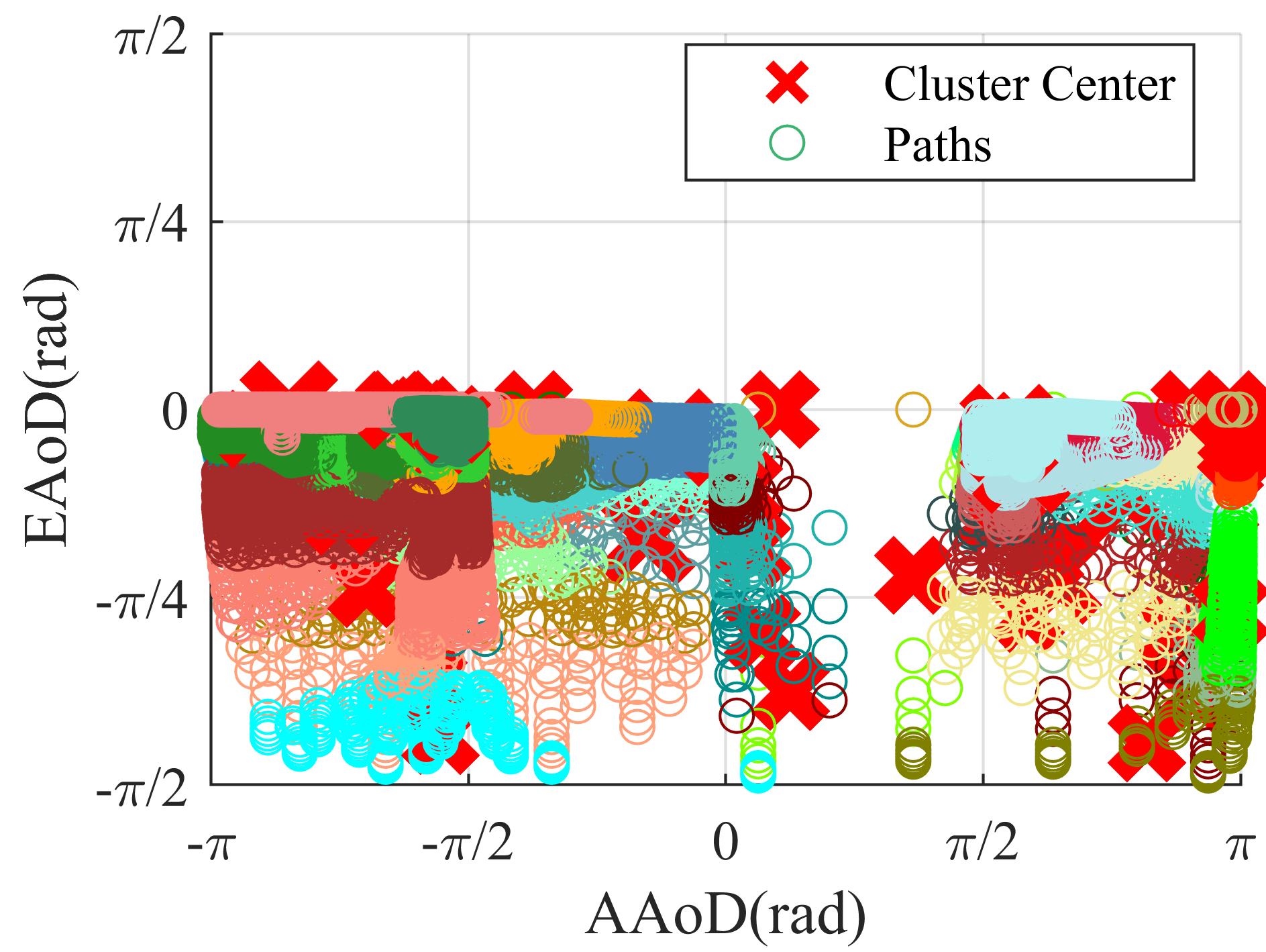}
		\label{fig:indoor2dcluster}
	}
	\hfill
	\subfloat[Angle-delay domain.]{
		\includegraphics[width=0.6\linewidth]{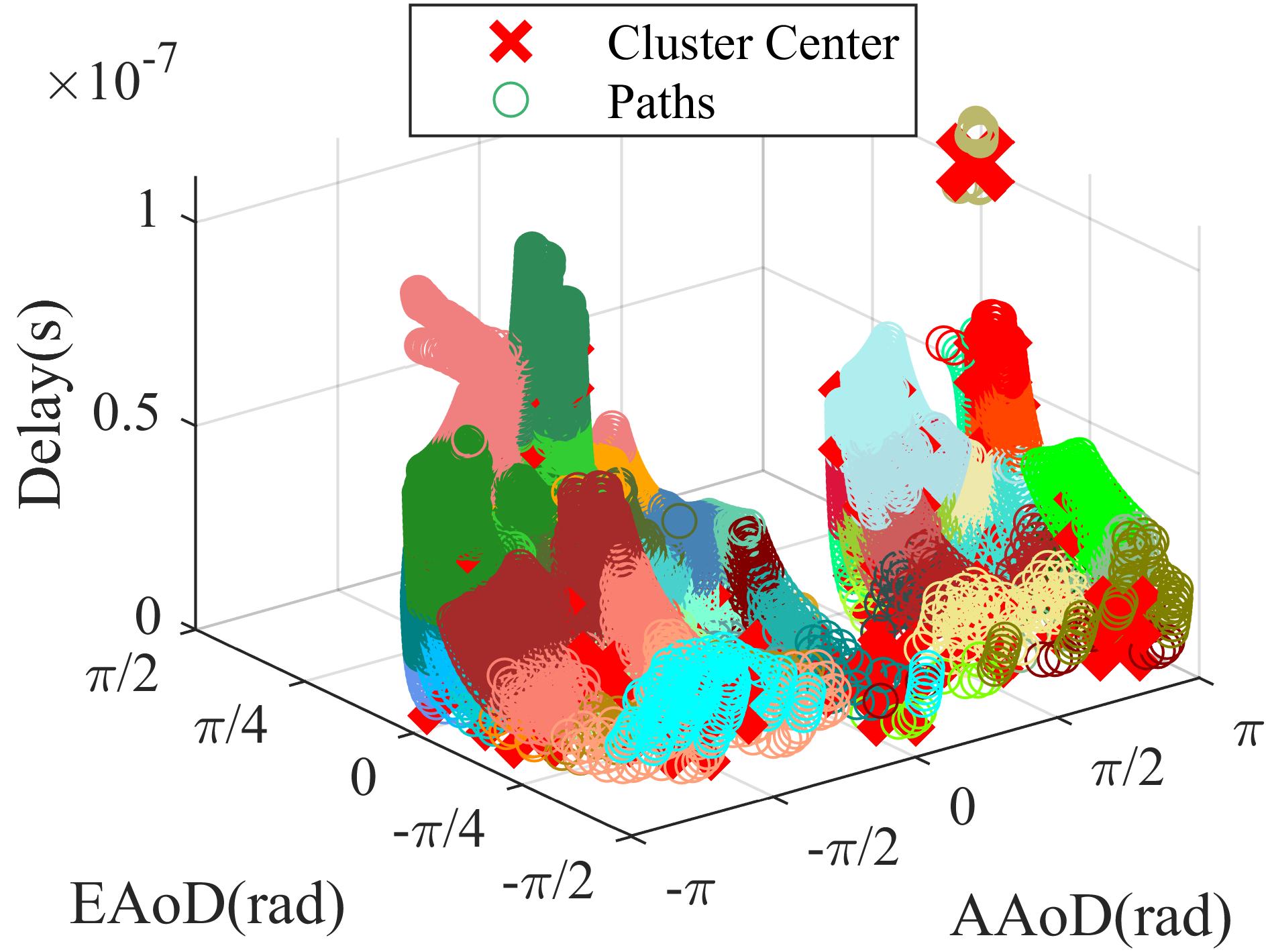}
		\label{fig:indoor3dcluster}
	}
	\hfill
	\subfloat[One cluster crossing $\pi$ in AAoD domain.]{
		\centering
		\includegraphics[width=0.6\linewidth]{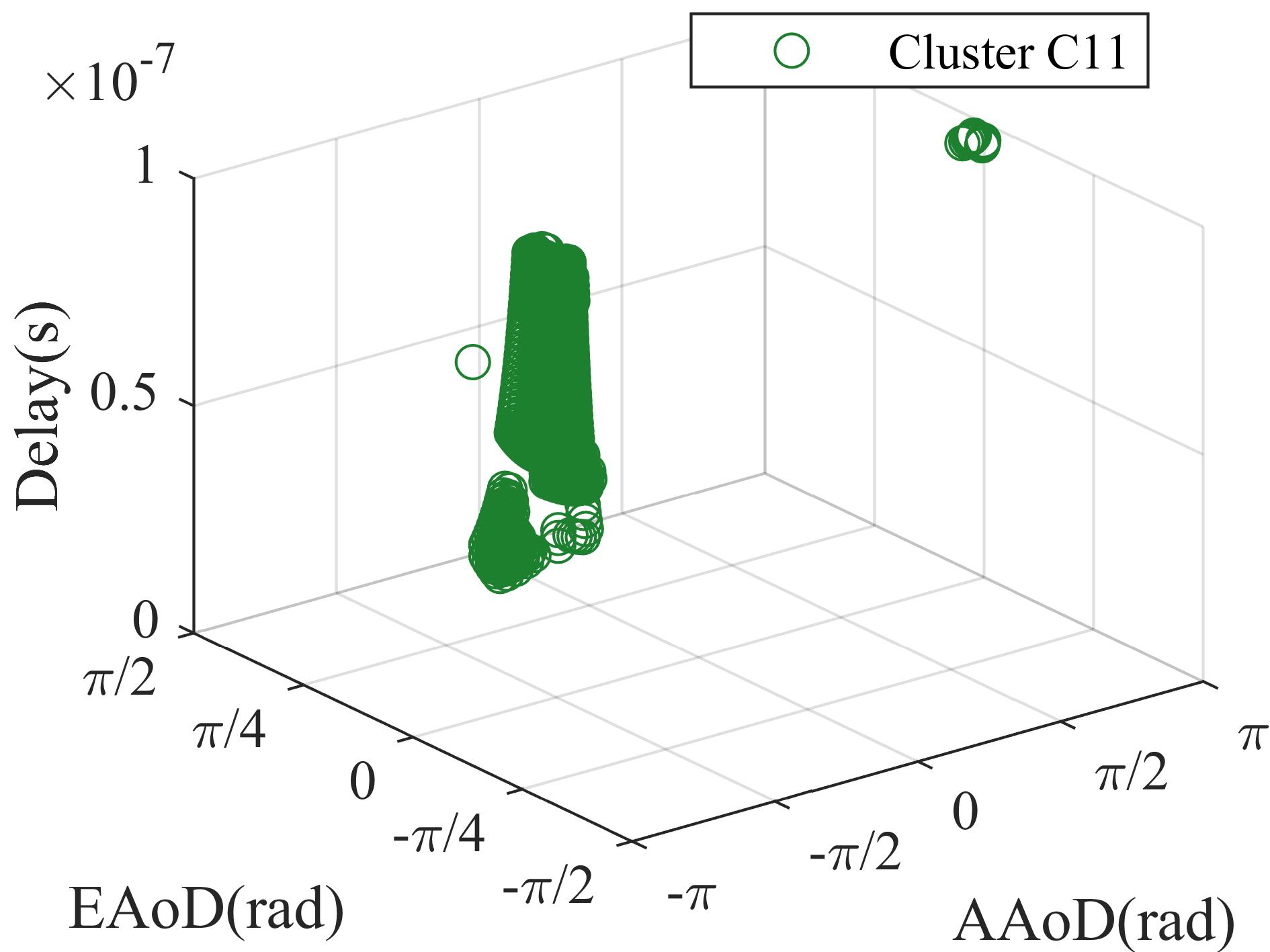}
		\label{fig:ACindoor}
	}
	\hfill
	\caption{Clusters in different shapes and colors.}
	\label{fig:indoor28clusters}
\end{figure}

Fig.~\ref{fig:indoor28clusters} demonstrates the clusters in the indoor 28~GHz scenario in AoD domain and angle-delay domain. Clusters in Fig.~\ref{fig:indoor28clusters}(a) seriously overlap each other in AoD domain, because existent paths with similar AAoD and EAoD, and diverse delays, are assigned to separate clusters, as shown in Fig.~\ref{fig:indoor28clusters}(b). Fig.~\ref{fig:indoor28clusters}(c) presents an example of grouping existent paths distributed near $\pi$ in AAoD domain into one cluster ($C_{11}$), proving that the proposed FCM addresses the angle periodicity problem in the conventional FCM. 
Besides, the minimum RSS among all existent paths is $-62.23$~dBm while the minimum cluster power is $-58.26$~dBm. In this case, UTs having the existent path with low RSS may receive a signal via the corresponding cluster with power level below the receiver sensitivity if the antenna gain in the BS is designed tightly in terms of the cluster power. Thus, the difference between the cluster power and the RSS of existent paths should be taken into consideration to meet the link budget when designing the antenna gain based on the proposed channel model.

\begin{figure}[t!]
	\centering
	\includegraphics[width=0.6\linewidth]{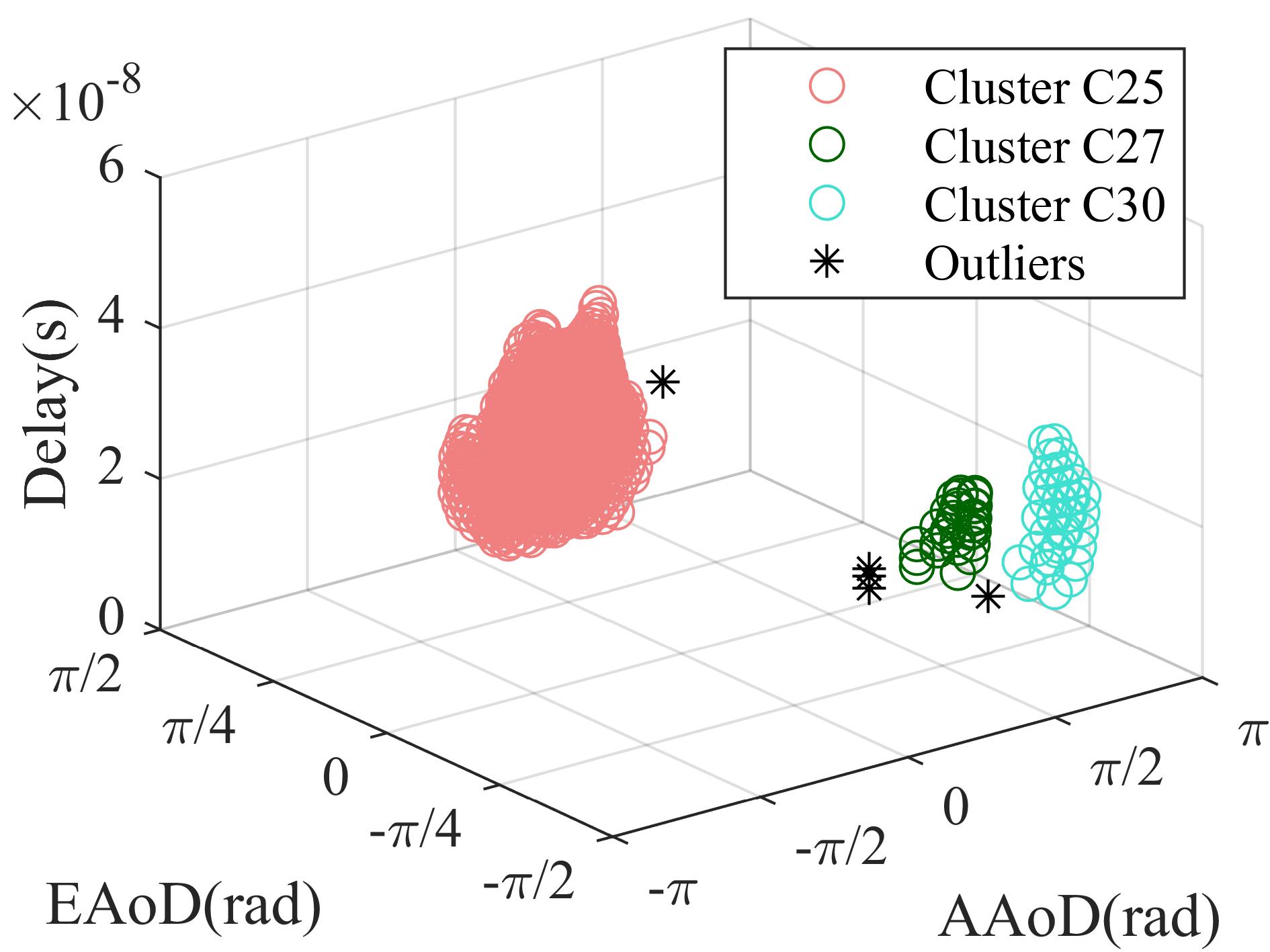}
	\caption{Pruned clusters in angle-delay domain.}
	\label{fig:mpcpruning}
\end{figure}

After the path pruning, the number of clusters decreases to $N_c=56$ in order to reduce the channel complexity. Fig.~\ref{fig:mpcpruning} presents three pruned clusters: $C_{25}$, $C_{27}$, $C_{30}$ in the angle-delay domain, where paths that are independent from the dense paths in the cluster are successfully identified and removed by the path pruning algorithm. The cluster parameters of $C_{25}$ are almost consistent before and after the path pruning while that of $C_{27}$ has a large drop in AAoD spread, from $0.3266$ rad to $0.1227$ rad, as summarized in Table~\ref{tab:pruningLSP}. It is because dropped outliers in $C_{27}$ are spatially distant from its cluster center centroid, which falsefully shifts the cluster center centroid in the direction with sparse paths.

\begin{table*}[t!]
	\centering
	\caption{Cluster Parameters before and after the Path Pruning.}
	\label{tab:pruningLSP}
	\begin{tabular}{|l|l|l|l|l|l|l|l|l|} 
	\hline
	Cluster                    & Param.                  & Pruning & AAoD(rad) & EAoD(rad) & AAoA(rad) & EAoA(rad) & RSS(dB)  & Delay($\mu s$)   \\ 
	\hline
	\multirow{4}{*}{$C_{25}$ } & \multirow{2}{*}{Mean}   & Before  & -1.9018   & -0.1819   & -1.9190   & -0.1819   & -81.9265 & 0.03299          \\ 
	\cline{3-9}
							   &                         & After   & -1.9020   & -0.1819   & -1.9189   & -0.1819   & -81.9240 & 0.03298          \\ 
	\cline{2-9}
							   & \multirow{2}{*}{Spread} & Before  & 0.3261    & 0.0477    & 0.3070    & 0.0477    & 6.3356   & 0.005914         \\ 
	\cline{3-9}
							   &                         & After   & 0.3257    & 0.0477    & 0.3069    & 0.0477    & 6.3377   & 0.005914         \\ 
	\hline
	\multirow{4}{*}{$C_{27}$ } & \multirow{2}{*}{Mean}   & Before  & 2.8342    & 0         & 2.8342    & 0         & -63.6223 & 0.002064         \\ 
	\cline{3-9}
							   &                         & After   & 2.9367    & 0         & 2.9367    & 0         & -64.2163 & 0.002030         \\ 
	\cline{2-9}
							   & \multirow{2}{*}{Spread} & Before  & 0.3265    & 0         & 0.3265    & 0         & 7.6060   & 0.002081         \\ 
	\cline{3-9}
							   &                         & After   & 0.1227    & 0         & 0.1227    & 0         & 7.4862   & 0.002201         \\
	\hline
	\end{tabular}
	\end{table*}

\begin{figure}[t!]
	\centering
	\subfloat[Selected clusters.]{
		\includegraphics[width=0.6\linewidth]{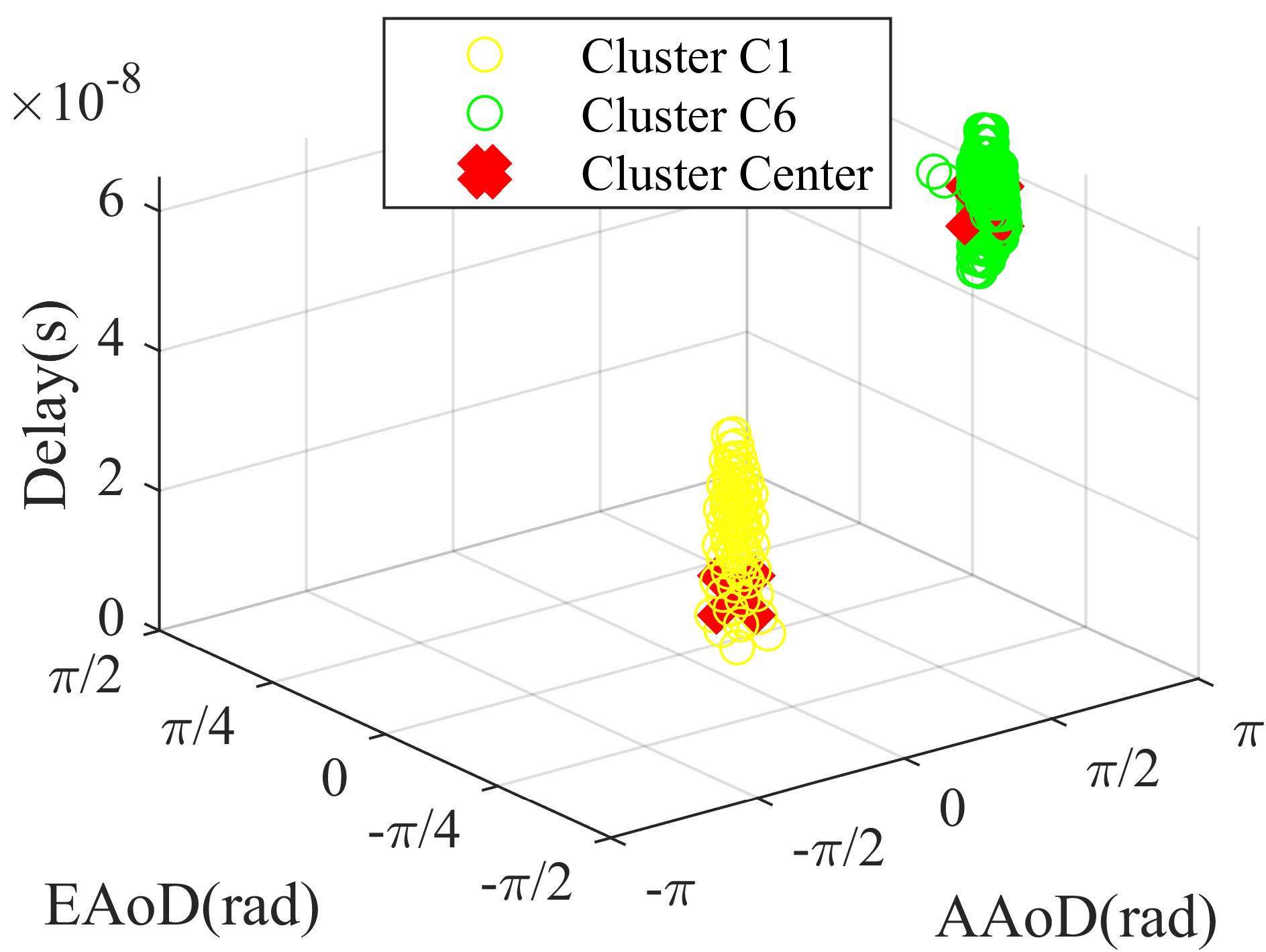}
	}
	\hfill
	\subfloat[Corresponding clusters regions.]{
		\includegraphics[width=0.6\linewidth]{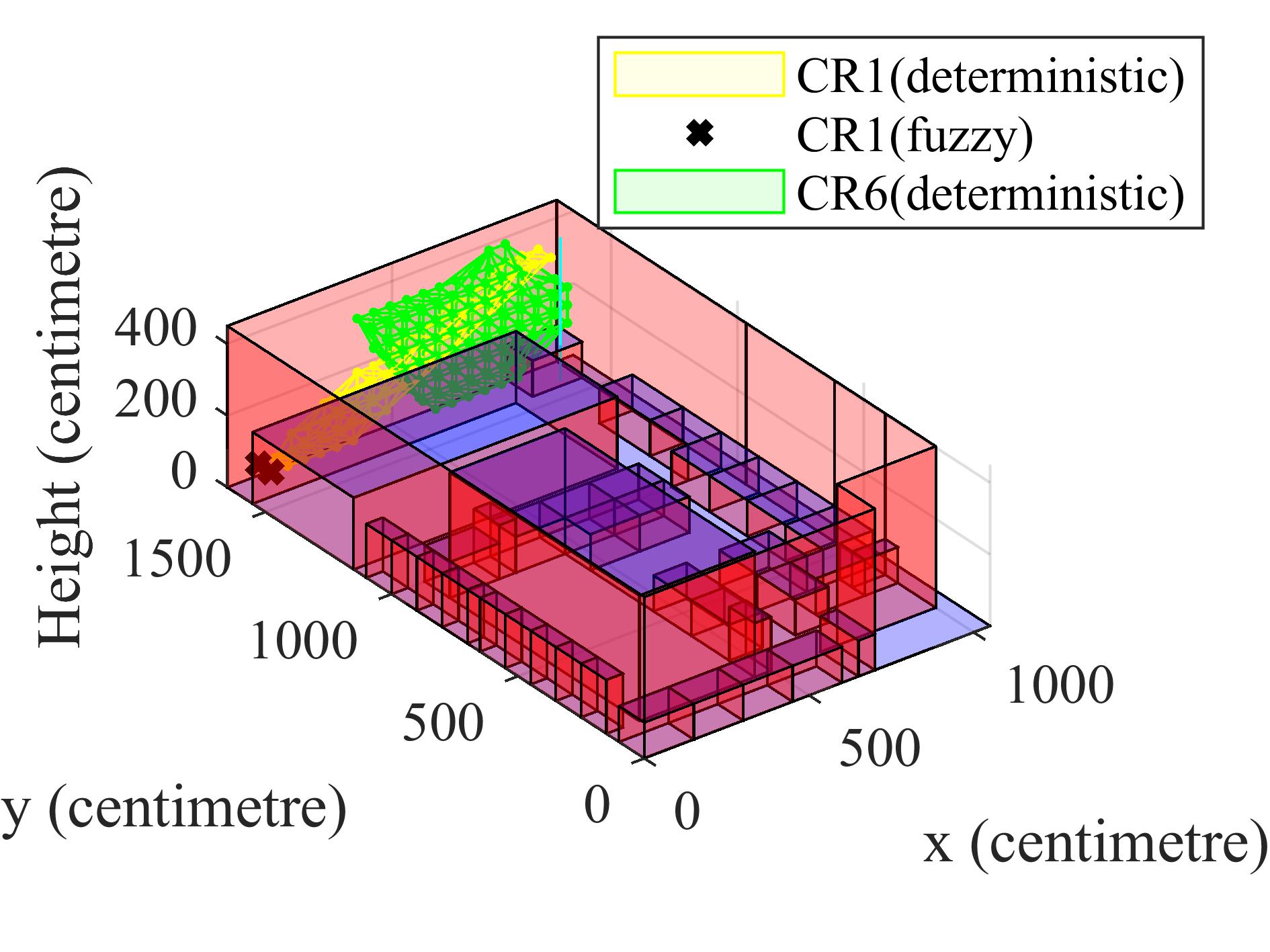}
	}
	\caption{Selected cluster regions.}
	\label{fig:IndoorVR}
\end{figure}

In Fig.~\ref{fig:IndoorVR}, two clusters ($C_1$ and $C_6$) result in an overlap in their cluster regions (CR1 and CR6). UTs located in the overlapping cluster region can receive signals through both clusters.

\begin{figure}[t!]
	\centering
	\subfloat[Low fuzziness.]{
		\includegraphics[width=0.6\linewidth]{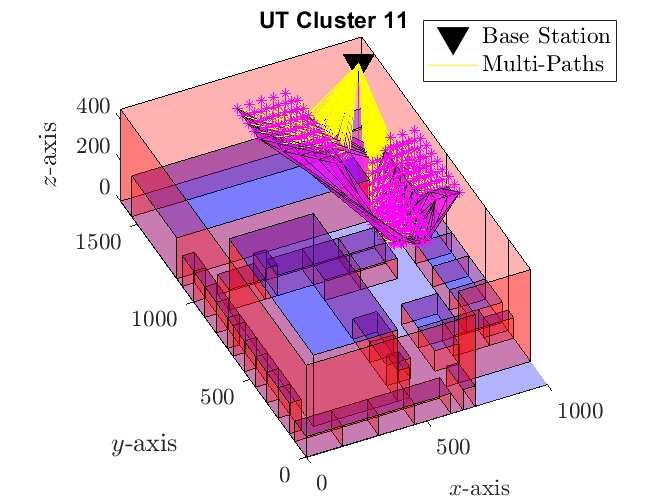}
	}
	\hfill
	\subfloat[Medium fuzziness.]{
		\includegraphics[width=0.6\linewidth]{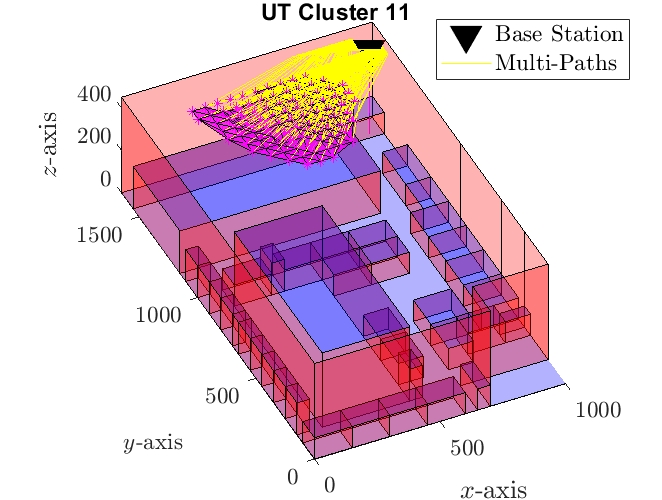}
	}
	\hfill
	\subfloat[High fuzziness.]{
		\includegraphics[width=0.6\linewidth]{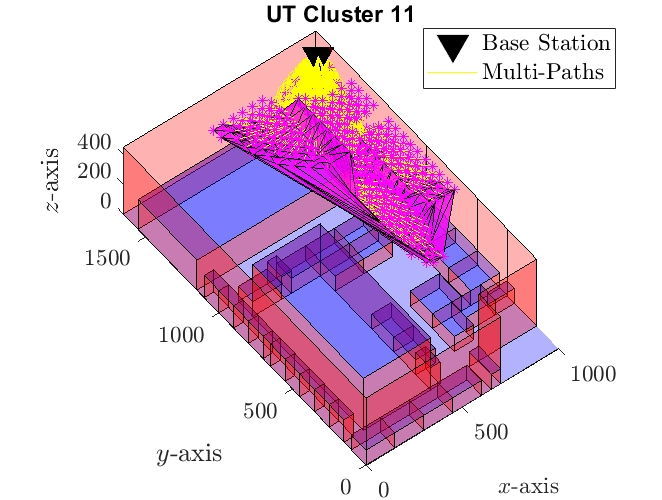}
	}
	\hfill
	\caption{Cluster region CR11 in different fuzziness levels.}
	\label{fig:fuzziness}
\end{figure}

Fig.~\ref{fig:fuzziness} shows that when the degree of the cluster fuzziness rises, existent paths in one cluster have more divergent AoD and decreasing average maximum membership (from $0.96$ to $0.025$), leading to less representative cluster parameters. Thus, the cluster fuzziness should be controlled at a low level.

\begin{figure}[t!]
	\centering
	\subfloat[UT zone map.]{
		\includegraphics[width=0.6\linewidth]{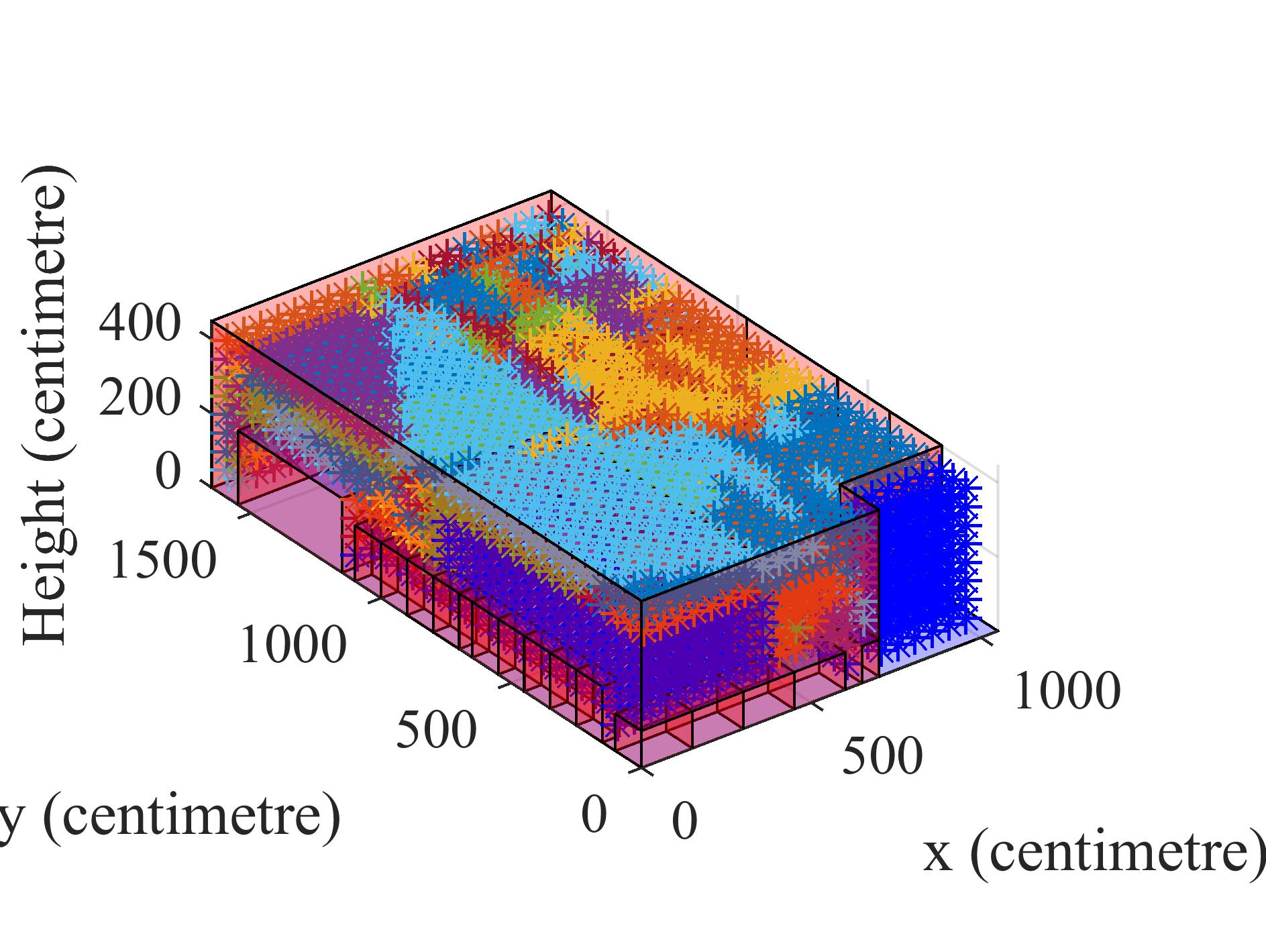}
	}
	\hfill
	\subfloat[Selected UT zones.]{
		\includegraphics[width=0.6\linewidth]{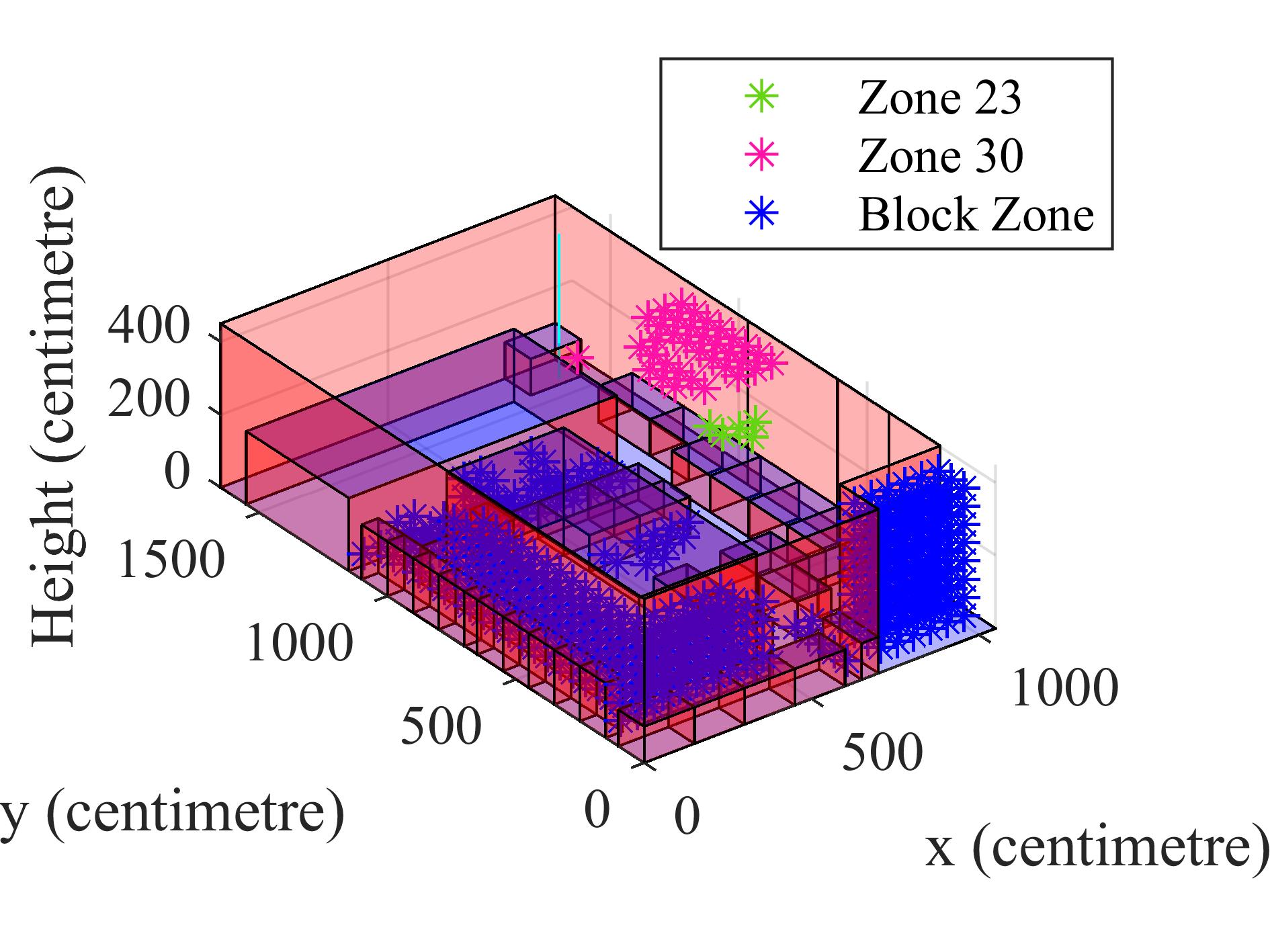}
	}
	\caption{UT zones for the indoor 28~GHz scenario.}
	\label{fig:AR}
\end{figure}

Fig.~\ref{fig:AR}(a) shows all UT zones for the indoor 28~GHz scenarios. Fig.~\ref{fig:AR}(b) depicts three zones, where Zone 23 is defined by the cluster group $\mathcal{Z}_{23}=\left\{C_{10}, C_{11}, C_{15}\right\}$, Zone 30 has $\mathcal{Z}_{30}=\left\{C_{10}, C_{11}, C_{35}\right\}$ and the block zone in blue has no cluster (out of transmission coverage). UTs in Zone 23 can pratically achieve maximum RSS by configuring RF beamformers with UT-group CSI of $\mathcal{Z}_{23}$, while the interference between UT locations in Zone 23 and Zone 30 can be cancelled by avoiding configuring with CSI of identical clusters ($C_{15}$ and $C_{35}$).

\section{Path Estimation based on DNN and 3D Geospatial Data}
The path estimation in Fig.~\ref{fig:Overview} spends long simulation/ measurement time to obtain UT-level CSI in terms of all paths required for the channel estimation completely by offline ray tracing. We design a DNN-based path estimation model to produce part of required paths for the channel estimation based on the geospatial data without ray tracing.
In this section, we study and develop DNN-based path estimation models, including a one-step model based on the FFNN and three two-step models based on the 1D-CNN, via three steps: (i) extract input features from the 3D geospatial data by data preprocessing, (ii) learn a small number of true paths of selected feasible UT locations and corresponding input features in the training set (generated by the offline ray tracing algorithm) by a DNN-based model, (iii) predict all paths for the remaining feasible UT locations only according to geospatial data by the trained DNN-based model.

\begin{figure}[t!]
	\centering
	\subfloat[Training phase.]{
		\includegraphics[width=\linewidth]{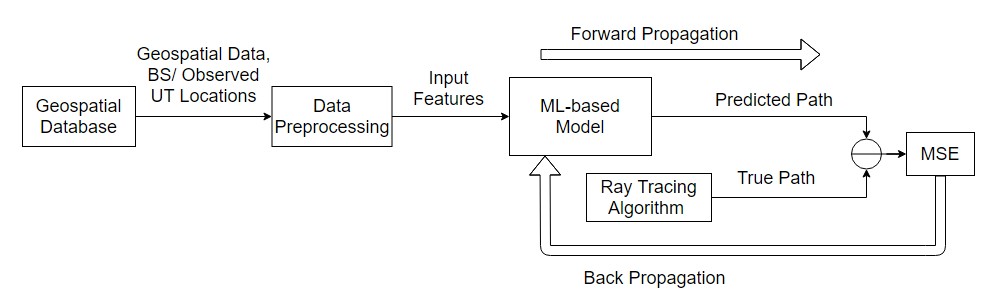}
		\label{fig:training}
	}
	\hfill
	\subfloat[Prediction phase.]{
		\includegraphics[width=\linewidth]{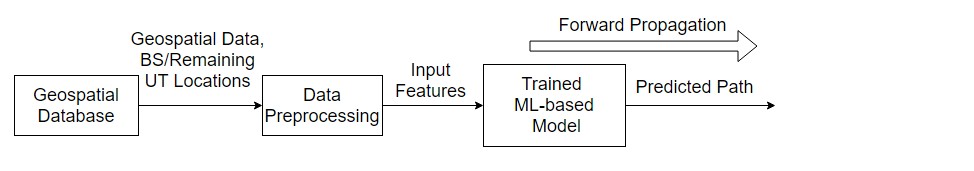}
		\label{fig:prediction}
	}
	\hfill
	\caption{DNN-based path estimation technique.}
	\label{fig:mlbasedapproach}
	\vspace{-2ex}
\end{figure}

\subsection{Data Preprocessing}
For the $u^{th}$ path, there are $15$ input features defined by $\mathcal{F}^u=\left\{\mathrm{V}_{BS}^u,\mathrm{V}_{UT}^u,\mathrm{V}_{\mathcal{P},1}^u,\mathrm{V}_{\mathcal{P},2}^u,\mathrm{V}_{\mathcal{P},3}^u\right\}$, where $\mathrm{V}_{BS}^u$ is the $x,y,z$ coordinates of the BS location, $\mathrm{V}_{UT}^u$ is the $x,y,z$ coordinates of the UT location, $\mathrm{V}_{\mathcal{P},1}^u$, $\mathrm{V}_{\mathcal{P},2}^u$, and $\mathrm{V}_{\mathcal{P},3}^u$ are the $x,y,z$ coordinates of three points $\mathrm{V_1,V_2,V_3}$ defining the interacted plane $\mathcal{P}$.

Consider a dataset $\mathbf{S}=\left\{\bm{o}_1,\cdots,\bm{o}_N\right\}$ with a total of $N$ paths for $K$ feasible UT locations required to comprehensively describe the desired channel, where $K$ feasible UT locations in the service area are known from the geospatial database, $\bm{o}_u$ is the vector composed of input features, the class tag and path parameters for the $u^{th}$ path. We build two groups according to $K$ feasible UT locations: one group with $K_{Tr}$ UT locations and the other group with $K_{Test}$ UT locations, where $K_{Tr}+K_{Test}=K$. Paths of $K_{Tr}$ UT locations are obtained by the offline ray tracing and taken as the training set $\mathbf{S}_{Tr}=\left\{\bm{o}_1,\cdots,\bm{o}_{N_{Tr}}\right\}$. Paths of $K_{Test}$ UT locations in the test set $\mathbf{S}_{Test}=\left\{\bm{o}_1,\cdots,\bm{o}_{N_{Test}}\right\}$ are unknown and required to be predicted by the proposed DNN-based path estimation model. To investigate the effects of training time on the prediction performance of the DNN-based path estimation model, we consider three sets of ratios for training and testing: $K_{Tr}:K_{Test}=0.7:0.3$, $K_{Tr}:K_{Test}=0.5:0.5$ and $K_{Tr}:K_{Test}=0.3:0.7$. If $K_{Tr}$ UT locations are densely distributed in the service area, the predicted paths for the remaining $K_{Test}$ UT locations would not well match their true paths. Thus, we assume that $K_{Tr}$ UT locations are randomly and uniformly selected from $K$ feasible UT locations without replacement.

\subsection{One-step Model}
\label{sec:onestep}
\begin{figure}[htbp]
	\centering
	\includegraphics[width=\linewidth]{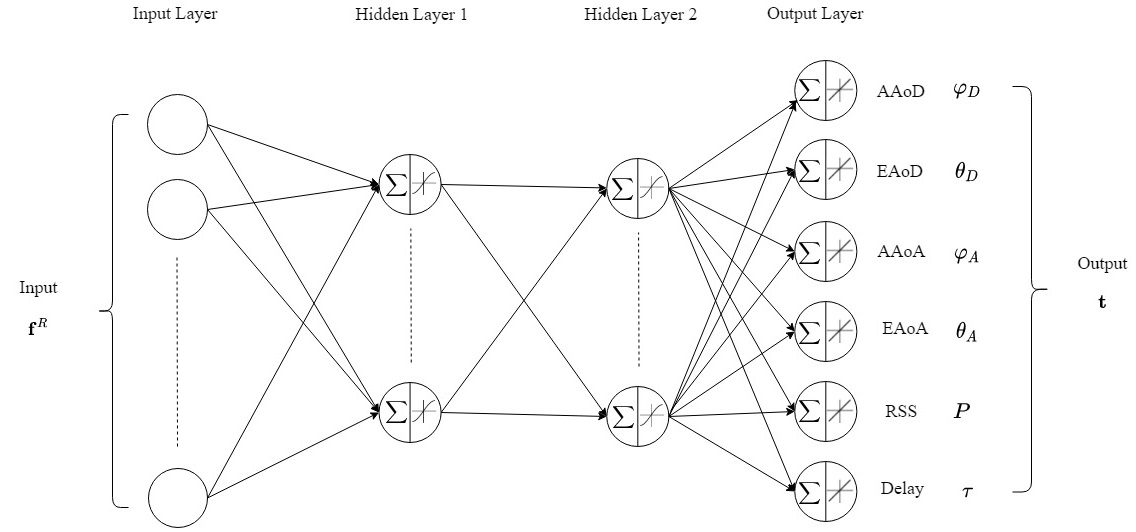}
	\caption{Structure of one-step model.}
	\label{fig:FFNN}
\end{figure}
The one-step model maps the input features to the path parameters by a FFNN as shown in Fig.~\ref{fig:FFNN}, {where the hidden layer is activated by the $tanh$ function and the output layer is activated by the linear function.} For the $u^{th}$ path, 15 input features $\mathcal{F}^u$ form an input vector $\mathbf{f}_u^R\in \mathbb{R}^{1\times 15}$, and six path parameters form an output vector $\mathbf{t}_u\in \mathbb{R}^{1\times 6}$. We assume that the topology of the one-step model that fits the mapping from the input vector to one of the six path parameters (i.e., AAoD) also fits the mapping from the input vector to all six path parameters. Hence, a number of FFNNs in different topologies are trained with the input vector and AAoD (i.e., $(\mathbf{f}_u^R, \varphi_{D_u})$ for the $u^{th}$ path) in the training set by Levenberg-Marquardt backpropagation algorithm, and evaluated by the root mean square error (RMSE) and R. We choose the topology of the one-step model, such as the number of hidden layers and the number of neurons in each hidden layer, from the FFNN with a lower RMSE and a higher R. 
Afterwards, the coefficients of the one-step model are obtained by training it with paths in $\mathbf{S}_{Tr}$ (i.e., $(\mathbf{f}_u^R, \mathbf{t}_u)$ for the $u^{th}$ path).

\subsection{Two-step Model}
\label{sec:twostep}
\begin{figure}[htbp]
	\centering
	\includegraphics[width=\linewidth]{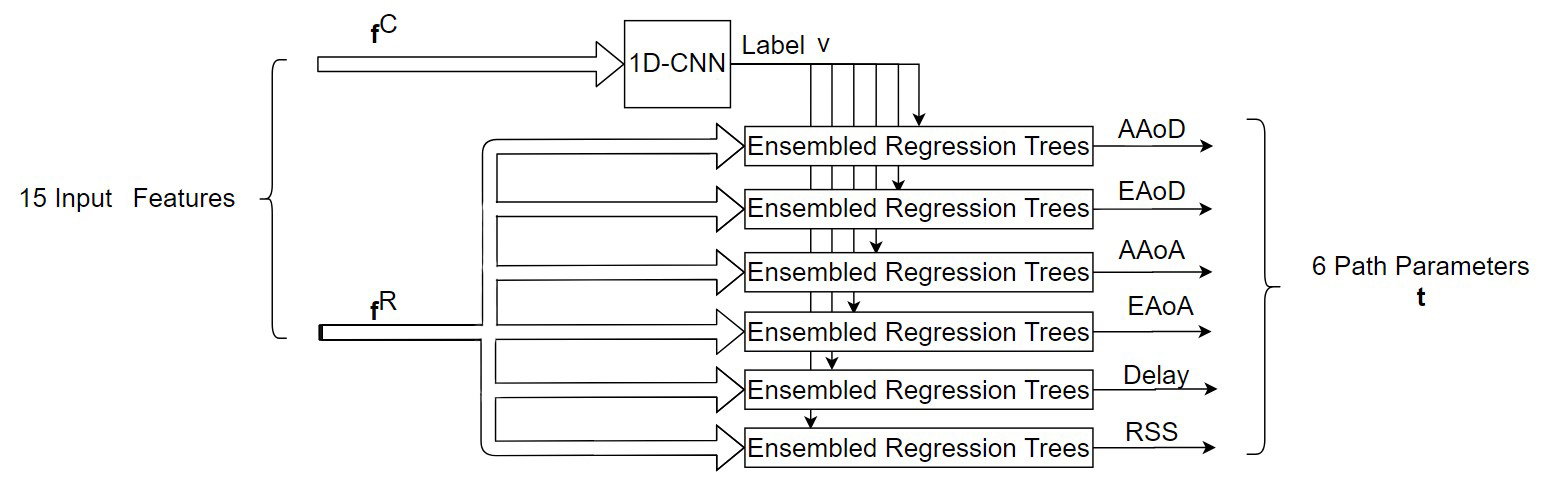}
	\caption{Structure of two-step model.}
	\label{fig:NN}
\end{figure}
The 1D-CNN and the regression tree ensembles are used in the proposed two-step model as presented in Fig.~\ref{fig:NN}, where the 1D-CNN classifies a large number of paths into two classes: the existent class and the non-existent class, and the regression tree ensembles only apply to a reduced small number of existent paths for path parameter prediction. The reasons why the 1D-CNN and the regression tree emsemble win against other machine learning techniques are that the 1D-CNN can preserve the relative relationships between $x, y, z$ coordinates in the input features compared to the FFNN in the one-step model, and the regression tree ensemble can handle small datasets with a high degree of errors and missing values with balanced variance and bias.

\subsubsection{One-dimensional convolutional neural network}
\begin{figure}[t!]
	\centering
	\includegraphics[width=\linewidth]{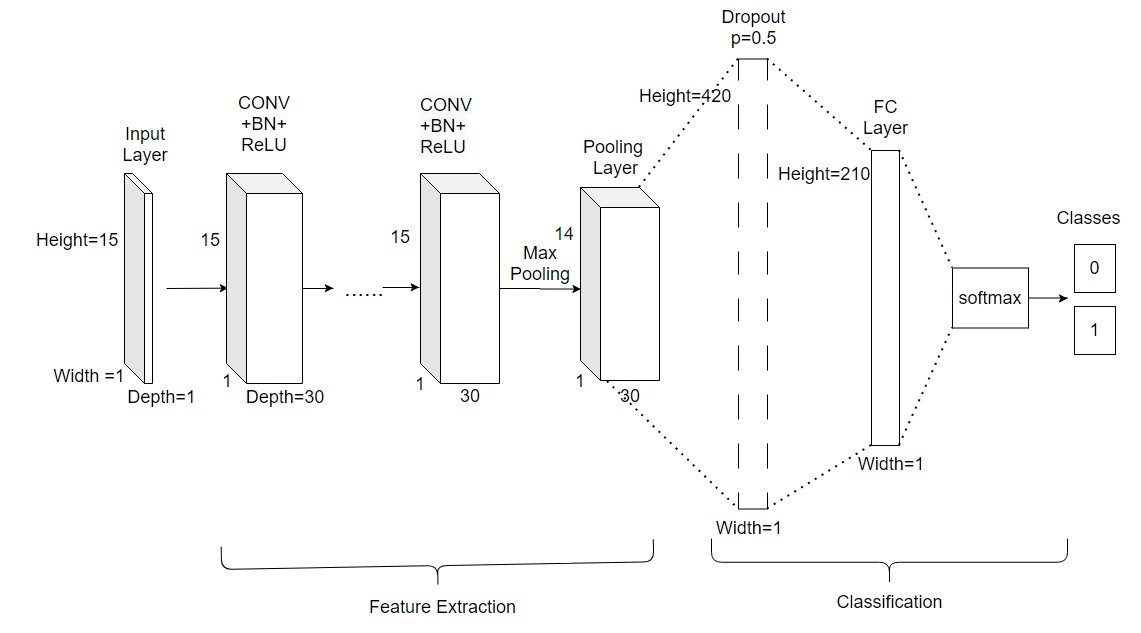}
	\caption{Structure of 1D-CNN.}
	\label{fig:CNN}
\end{figure}
For the $u^{th}$ path, 15 input features $\mathcal{F}^u$ are organized as a three-dimensional input vector $\mathbf{f}_u^{c}\in \mathbb{R}^{1 \times 15\times 1}$, {and the corresponding output is the categorical class tag $\mathrm{v}_u$ with two values,} where $\mathrm{v}_u=1$ if the $u^{th}$ path is existent and $\mathrm{v}_u=0$ if the $u^{th}$ path is non-existent.

The proposed 1D-CNN is composed of an input layer, $\mathrm{L_c}$ convolutional (CONV) layers, a max pooling layer, a dropout layer and a fully connected (FC) layer, as shown in Fig.~\ref{fig:CNN}. In the $\mathrm{l}^{th}$ CONV layer, the input vector $\mathbf{I}_{\mathrm{l}} \in \mathbb{R}^\mathrm{15\times 1\times C_{il}}$, where $\mathrm{C_{il}}$ is the depth of the input vector, is normalized by the batch normalization technique and padded with zeros in the width and height dimensions for all depths $\mathbf{I}_{\mathrm{l}}^\mathrm{p} \in \mathbb{R}^\mathrm{17\times 3\times C_{il}}$, convolves with $\mathrm{C_{ol}}$ different filters $\bm{F}_\mathrm{j}\in \mathbb{R}^{3\times 3 \times \mathrm{C_{il}}},\mathrm{j}=1,\cdots,\mathrm{C_{ol}}$ (e.g., $\mathrm{C_{ol}}$=30) using stride $s=1$ to produce $\mathbf{I}_{\mathrm{l},j}^\mathrm{c} =\mathbf{I}_{\mathrm{l}}^\mathrm{p} \ast \bm{F}_\mathrm{j}\in \mathbb{R}^\mathrm{15\times 1\times 1}$. The output vector of the convolution operation $\mathbf{I}_{\mathrm{l}}^\mathrm{c} \in\mathbb{R}^\mathrm{15\times 1\times \mathrm{C_{ol}}}$ is activated by the rectified linear unit (ReLU) $\phi_\mathrm{ReLU}(\mathbf{I}_{\mathrm{l}}^\mathrm{c})$. The output vector of the $\mathrm{l}^{th}$ CONV layer is given by $\mathbf{O}_{\mathrm{l}}^{\mathrm{c}}=\phi_\mathrm{ReLU}(\mathbf{I}_{\mathrm{l}}^\mathrm{c})\in \mathbb{R}^\mathrm{15\times 1\times C_{ol}}$. After $\mathrm{L_c}$ consecutive CONV layers, the input vector of the following max pooling layer $\mathbf{I}^{\mathrm{P}}\in \mathbb{R}^\mathrm{15\times 1\times C_{oL_c}}$ is operated with a window of size $1\times 2$ and stride $s=1$, in order to extract the max value between the neighboring elements in each depth of the input vector. The output of the max pooling layer $\mathbf{O}^{\mathrm{P}}\in \mathbb{R}^\mathrm{14\times 1\times C_{oL_c}}$ is the input of the dropout layer with a dropout ratio of 0.5 to reduce the number of trainable parameters in the subsequent layers. The output of the dropout layer $\mathbf{O}^{\mathrm{D}}\in \mathbb{R}^\mathrm{7C_{oL_c}\times 1}$ connects to $N_i^{\mathrm{f}}=7C_{oL_c}$ (e.g., $210$) neurons in the FC layer. Finally, the FC layer activated by the softmax function maps $N_i^{\mathrm{f}}=7C_{oL_c}$ neurons to $N_o=2$ neurons to predict the class tag $\mathrm{\hat{v}}$ for each path.

As the number of paths in the training set for each class is imbalanced (existent:non-existent=2\%:98\%) as shown in Table~\ref{tab:indoor results}, the 1D-CNN is trained with the binary focal loss given by\cite{lin2017focal}:
\begin{equation}
	\mathrm{FL}= -\sum_{\mathrm{v}=0}^1(1 - p_{\mathrm{v}})^{\gamma }\alpha_{\mathrm{v}} log (p_{\mathrm{v}})
\end{equation}
where $p_{\mathrm{v}}$ is the probability of one path belonging to the class $\mathrm{v}$ such that $\sum_{\mathrm{v}=0}^1p_{\mathrm{v}}=1$, $\gamma$ is a modulating factor to weigh up the hard-classified paths, and $\alpha_\mathrm{v}$ is the weighting factor of the class $\mathrm{v}$ to weigh up the existent paths such that $\sum_{\mathrm{v}=0}^1\alpha_\mathrm{v}=1$.

The performance of the 1D-CNN is gauged by prediction, recall, F-score and Cohen's kappa instead of accuracy, since the class imbalance between the two classes leads to high accuracy by simply labeling all paths as non-existent paths\cite{valverde2014100}. 
Cohen's kappa $\kappa$ represents a minimum acceptable level of agreement of the predicted path class ranging from $0$ to $1$, where $\kappa=0$ denotes a slight agreement and $\kappa=1$ denotes an almost perfect agreement\cite{mchugh2012interrater}. It is given by:
\begin{equation}
	\label{eq:kappa}
	\kappa=\frac{p_0-p_e}{1-p_e},
\end{equation}
where $p_0$ is the proportion of observed agreement and $p_e$ is the proportion of chance agreement. For a binary-class problem, $p_e$ is given by\cite{mchugh2012interrater}:
\begin{equation}
	p_e=\frac{(TN+FN)(TN+FP)+(FP+TP)(TP+FN)}{N^2}.
\end{equation}

\subsubsection{Regression tree ensembles}
\begin{figure}[htbp]
	\centering
	\includegraphics[width=\linewidth]{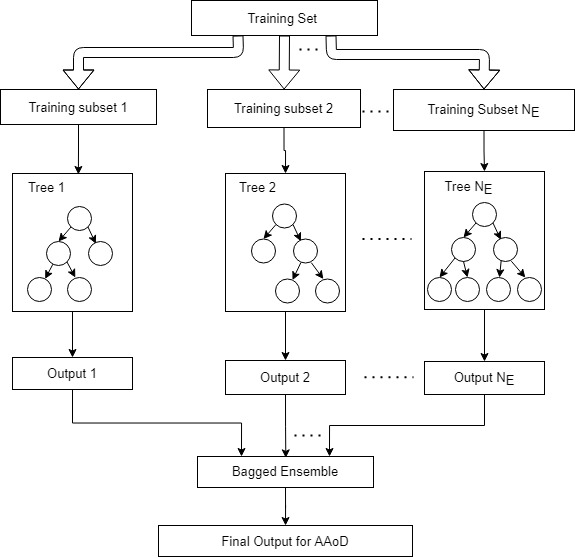}
	\caption{Structure of the regression tree ensemble for AAoD prediction.}
	\label{fig:Tree}
\end{figure}
There are six parallel regression tree ensembles, where each predicts one of the path parameters, as shown in Fig.~\ref{fig:Tree}. Unlike the training set $\mathrm{S}_{Tr}$ used in the 1D-CNN, this module is trained on the reduced training set $\mathbf{\tilde{S}}_{Tr}=\bm{[o}_1,\cdots,\bm{o}_{N_\mathrm{v}}]$, where only $N_\mathrm{v}$ out of $N_{Tr}$ paths in the existent class $\mathrm{v}=1$ are involved. For the $u^{th}$ existent path in $\mathbf{\tilde{S}}_{Tr}$, the input vector is $\mathbf{f}_u^R\in \mathbb{R}^{1\times 15}$ and the outputs are $\mathrm{t}_u=\left\{\bm{x}_u,P_u\right\}\in \mathbb{R}^{1 \times 6}$ for six regression tree ensembles, respectively. The performance of the regression tree ensemble is evaluated by RMSE.

For AAoD prediction, the entire training set of the regression tree ensemble is $\mathbf{\tilde{S}}_{Tr}^{D}$ (i.e., $(\mathbf{f}_u^R, \varphi_{D_u})$ for the $u^{th}$ existent path). There are $N_E$ regression trees in the regression tree ensemble. For the $e^{th}$ regression tree in the regression tree ensemble, its training subset $\mathbf{\tilde{S}}_{Tr}^{D_e}$ of size $N_e$ is generated from $\mathbf{\tilde{S}}_{Tr}^{D}$ of size $N_\mathrm{v}$ by randomly drawing $N_e$ paths paths with replacement. $\mathbf{\tilde{S}}_{Tr}^{D_e}$ is randomly divided into 5 folds for 5-fold cross validation. The output of the regression tree ensemble is produced by averaging the outputs of $N_E$ regression trees. 
In order to reduce the depth of regression trees, only 12 input features $\left\{\mathrm{V}_{UT}^u, \mathrm{V}_{\mathcal{P},1}^u, \mathrm{V}_{\mathcal{P},2}^u, \mathrm{V}_{\mathcal{P},3}^u\right\}$ are taken as candidate predictors, as $\mathrm{V}_{BS}^u$ is the same for all paths in $\mathbf{\tilde{S}}_{Tr}$.

\subsection{Illustrative Results}
To evaluate the performance of the proposed DNN-based path estimation framework, we present a one-step model trained with 70\% of paths and three two-step models trained with 70\%, 50\% and 30\% of paths required for the channel estimation for the indoor 28~GHz scenario. The search range of the hyperparameters for the DNN-based path estimation model is given in Table~\ref{tab:CNN Hyper}.

\begin{table}[t!]
	\centering
	\caption{Hyperparameter Search Range}
	\label{tab:CNN Hyper}
	\begin{tabular}{|l|l|l|} 
	\hline
	Module                                                                              & Hyperparameters                & Search Range           \\ 
	\hline
	\multirow{5}{*}{FFNN}                                                               & Number of hidden layers        & 1-2                    \\ 
	\cline{2-3}
																						& Number of hidden neurons       & 10-30                  \\ 
	\cline{2-3}
																						& Training algorithm              & Levenberg-Marquardt    \\ 
	\cline{2-3}
																						& Damping factor                 & $10^{-6}$ - 10  \\ 
	\cline{2-3}
																						& Batch size                     & 1                      \\ 
	\hline
	\multirow{5}{*}{1D-CNN}                                                             & Number of CONV layers          & 1-12                   \\ 
	\cline{2-3}
																						& Number of Filters              & 10-50                  \\ 
	\cline{2-3}
																						& $\alpha_1$ in FL               & 0.1-0.4                \\
																						\cline{2-3}
																						& Training  algorithm                & RMSProp   \\ 
	\cline{2-3}
																						& Learning rate                  & $10^{-8}$-0.001        \\ 
	\cline{2-3}
																						& Batch size                     & 4096                   \\ 
	\hline
	\multirow{3}{*}{\begin{tabular}[c]{@{}l@{}}Regression\\Tree\\Ensemble\end{tabular}} & Minimum leaf size              & 1-6924                 \\ 
	\cline{2-3}
																						& Number of learners             & 10-500                 \\ 
	\cline{2-3}
																						& Number of predictors to sample & 1-12                   \\
	\hline
	\end{tabular}
\end{table}

\subsubsection{One-step model}
\begin{figure}[t!]
	\centering
	\includegraphics[width=0.5\linewidth]{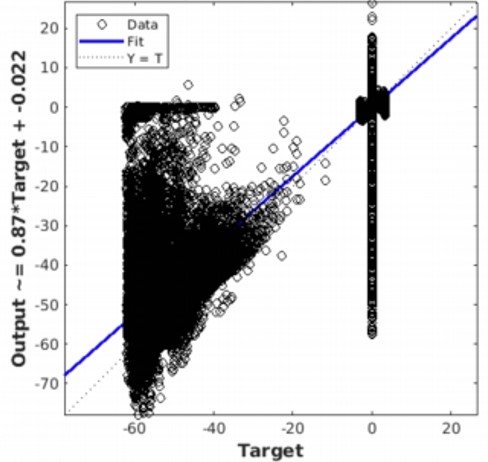}
	\caption{Fitting curve of the best topology for AAoD prediction.}
	\label{fig:FFNNAoD}
\end{figure}

The performance of the one-step model for AAoD prediction trained with 70\% of paths is shown in Table~\ref{tab:FFNNresults}. The best topology of the FFNN for AAoD prediction has $25$ neurons in the first hidden layer and $15$ neurons in the second hidden layer. It achieves $RMSE=0.0995$~rad and $R=0.924$ on the test set, indicating that this network learns well from the training samples. However, Fig.~\ref{fig:FFNNAoD} shows that there are huge differences between the predicted AAoD and the true AAoD for specific paths. Particularly, for a non-existent path, the one-step model falsefully assigns a non-zero value to its AAoD.

\begin{table}[t!]
	\centering
	\caption{Performance of the One-step Model with Different Topologies for AAoD Prediction.}
	\label{tab:FFNNresults}
	\begin{tabular}{|l|l|l|} 
	\hline
	Topology                          & RMSE on AAoD (rad) & R      \\ 
	\hline
	10                                & 0.1558             & 0.755  \\ 
	\hline
	15                                & 0.1600             & 0.620  \\ 
	\hline
	20                                & 0.1534             & 0.718  \\ 
	\hline
	30                                & 0.1401             & 0.781  \\ 
	\hline
	20/15                             & 0.1074             & 0.916  \\ 
	\hline
	25/10                             & 0.1059             & 0.912  \\ 
	\hline
	\rowcolor[rgb]{0.8,0.8,0.8} 25/15 & 0.0995             & 0.924  \\ 
	\hline
	25/20                             & 0.1129             & 0.912  \\
	\hline
	\end{tabular}
	\end{table}

\subsubsection{Two-step model trained with 70\% of required paths}
For the 1D-CNN, we first search for a best $\alpha_1$ in the focal loss, followed by the number of filters in each CONV layer and the number of CONV layers by training them with 70\% of required paths. Then we find the best topologies for six regression tree ensembles. Table~\ref{tab:CNN alpha} shows that two $\alpha_1$, $0.25$ and $0.32$ with the top two Cohen's kappa, are selected for further hyperparameter tuning.

\begin{table}[t!]
	\centering
	\caption{Performances of Shallow 1D-CNNs for $\alpha_1$ Ranging from $0.1$ to $0.4$}
	\label{tab:CNN alpha}
	\begin{tabular}{|l|l|l|l|l|} 
	\hline
	$\alpha_1$                       & Recall(\%) & Precision(\%) & F-score(\%) & Cohen's kappa  \\ 
	\hline
	0.1                              & 18.4533   & 93.5922      & 30.8283    & 0.3038         \\ 
	\hline
	0.2                              & 10.2603   & 97.8102      & 18.5720    & 0.1827         \\ 
	\hline
	\rowcolor[rgb]{0.8,0.8,0.8} 0.25 & 30.8576   & 86.5736      & 45.4980    & 0.4493         \\ 
	\hline
	0.3                              & 17.4962   & 95.2083      & 29.5602    & 0.2913         \\ 
	\hline
	0.31                             & 16.8836   & 96.0784      & 28.7202    & 0.2830         \\ 
	\hline
	\rowcolor[rgb]{0.8,0.8,0.8} 0.32 & 22.1669   & 91.3249      & 35.6747    & 0.3518         \\ 
	\hline
	0.33                             & 17.8407   & 95.1020      & 30.0451    & 0.2961         \\ 
	\hline
	0.35                             & 18.9510   & 83.8983      & 30.9182    & 0.3042         \\ 
	\hline
	0.36                             & 17.5727   & 94.0574      & 29.6129    & 0.2918         \\ 
	\hline
	0.37                             & 19.0276   & 92.2078      & 31.5455    & 0.3108         \\ 
	\hline
	0.4                              & 17.9173   & 95.3157      & 30.1640    & 0.2973         \\
	\hline
	\end{tabular}
	\end{table}

Table~\ref{tab:CNN focal loss} illustrates that the 1D-CNN trained using the focal loss with $\alpha_1=0.32$ show a great increase in precision (3\%), recall (0.7\%), F-score (1.8\%) and Cohen's kappa (2\%), compared with the 1D-CNN trained using the conventional cross entropy loss function. However, the model trained with the focal loss having $\alpha_1=0.1$ performs worse than models in other three settings. Thus, we set $\alpha_1=0.32$ in the focal loss function for the desired 1D-CNN.

\begin{table}[t!]
	\centering
	\caption{Performances of 5-CONV-layer 1D-CNNs with Focal Loss and Cross Entropy Loss}
	\label{tab:CNN focal loss}
	\begin{tabular}{|l|l|l|l|l|} 
	\hline
	 $\alpha_1$                      & Recall(\%) & Precision(\%) & F-score(\%) & Cohen's kappa  \\ 
	\hline
	-                                & 87.2894   & 92.4949      & 89.8168    & 0.8962         \\ 
	\hline
	0.1                              & 74.7703   & 94.9441      & 83.6582    & 0.8337         \\ 
	\hline
	0.25                             & 87.5957   & 95.6521      & 91.4468    & 0.9128         \\ 
	\hline
	\rowcolor[rgb]{0.8,0.8,0.8} 0.32 & 88.0934   & 95.4376      & 91.6146    & 0.9146         \\
	\hline
	\end{tabular}
	\end{table}

Table~\ref{tab:CNN filter} shows that the 1D-CNN with $30$ filters in all CONV layers performs in a superior way than 1D-CNNs with more or less filters in each CONV layer, with recal$l=88.09\%$, precision$=95.44\%$, F-score$=91.61\%$ and Cohen's kappa$=0.91$. Thus, we set $30$ filters in all CONV layers for the desired 1D-CNN.

\begin{table}[t!]
	\centering
	\caption{Performances of 5-CONV-layer 1D-CNNs with Different Number of Filters in CONV Layers}
	\label{tab:CNN filter}
	\begin{tabular}{|l|l|l|l|l|} 
		\hline
		\# of Filters  & Recall(\%) & Precision(\%) & F-score(\%) & \begin{tabular}[c]{@{}l@{}}Cohen's\\~kappa \end{tabular}  \\ \hline
		10/10/10/10/10  & 31.59      & 93.43   & 47.21    & 0.47  \\ \hline
		20/20/20/20/20  & 68.72      & 91.16   & 78.37    & 0.78   \\ \hline
		20/30/50/50/30  & 81.16      & 94.85   & 87.48    & 0.87   \\ \hline
		\rowcolor{gray!40}
		30/30/30/30/30  & 88.09      & 95.44   & 91.61    & 0.91  \\ \hline
		30/30/50/50/20  & 75.01      & 93.69   & 83.31    & 0.83    \\ \hline
		40/40/40/40/40  & 84.84      & 93.66   & 89.03    & 0.89     \\ \hline
		50/50/50/50/50  &  85.68    &  91.61   &  88.55   & 0.88    \\\hline
	\end{tabular}
\end{table}

\begin{figure}[t!]
	\centering
	\includegraphics[width=0.8\linewidth]{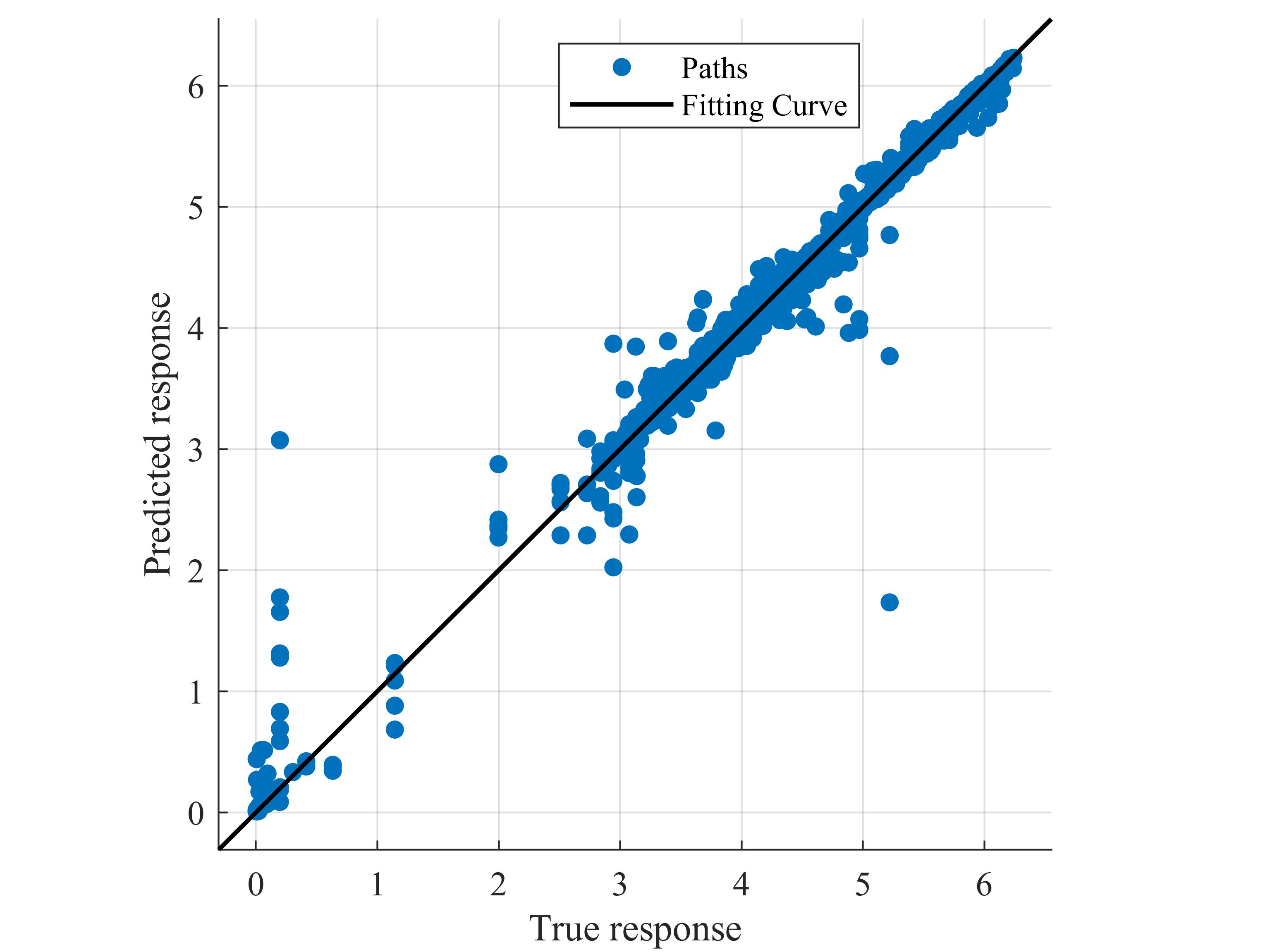}
	\caption{Predicted and true paths in AAoD prediction.}
	\label{fig:treeAoD}
\end{figure}

Table~\ref{tab:CNN layers} illustrates that \#8 1D-CNN with $9$ CONV layers with recall$=92.00\%$, precision$=96.31\%$, F-score$=94.11\%$ and Cohen's kappa$=0.9399$, leads to a perfect fit to the observed input-output pairs, compared to \#1 1D-CNN with 2 CONV layers which is too crude to perform path prediction due to its unacceptable Cohen's kappa ($0.29$) and compared to \#9 1D-CNN with $10$ CONV layers which has a worse performance. Thus, the two-step model trained with 70\% of required paths consist of \#8 1D-CNN and regression tree ensembles summarized in Table~\ref{tab:0.7RT}. Meanwhile, Fig.~\ref{fig:treeAoD} presents predicted and true paths in AAoD, where the predicted values closely approach the ideal fitting curve.

\begin{table}[t!]
	\centering
	\caption{Performances of 1D-CNNs Trained with 70\% of Required Paths}
	\label{tab:CNN layers}
	\begin{tabular}{|l|l|l|l|l|l|} 
	\hline
	\#                            & \begin{tabular}[c]{@{}l@{}}\# of \\CONV\\layers \end{tabular} & Recall  & Precision & F-score & \begin{tabular}[c]{@{}l@{}}Cohen's\\kappa \end{tabular}  \\ 
	\hline
	1                             & 2                                                             & 17.50\% & 95.21\%   & 29.56\% & 0.29                                                     \\ 
	\hline
	2                             & 3                                                             & 28.63\% & 90.34\%   & 43.49\% & 0.43                                                     \\ 
	\hline
	3                             & 4                                                             & 67.73\% & 90.63\%   & 77.52\% & 0.77                                                     \\ 
	\hline
	4                             & 5                                                             & 87.60\% & 95.65\%   & 91.45\% & 0.91                                                     \\ 
	\hline
	5                             & 6                                                             & 86.37\% & 95.47\%   & 90.69\% & 0.91                                                     \\ 
	\hline
	6                             & 7                                                             & 90.20\% & 95.93\%   & 92.98\% & 0.9284                                                   \\ 
	\hline
	7                             & 8                                                             & 92.00\% & 96.12\%   & 94.01\% & 0.9390                                                   \\ 
	\hline
	\rowcolor[rgb]{0.8,0.8,0.8} 8 & 9                                                             & 92.00\% & 96.31\%   & 94.11\% & 0.9399                                                   \\ 
	\hline
	9                             & 10                                                            & 92.00\% & 95.97\%   & 93.94\% & 0.9382                                                   \\
	\hline
	\end{tabular}
\end{table}

\begin{table}[t!]
	\centering
	\caption{Performances of Regression Tree Ensembles Trained with 70\% of Required Paths.}
	\label{tab:0.7RT}
	\begin{tabular}{|l|l|l|l|} 
	\hline
	Target                     & \multicolumn{2}{l|}{Optimized Hyperparameters} & Val. RMSE                \\ 
	\hline
	\multirow{3}{*}{AAoD(rad)} & Number of learners         & 11                & \multirow{3}{*}{0.0525}  \\ 
	\cline{2-3}
							   & Minimum leaf size          & 2                 &                          \\ 
	\cline{2-3}
							   & \# of predictors to sample & 12                &                          \\ 
	\hline
	\multirow{3}{*}{EAoD(rad)} & Number of learners         & 27                & \multirow{3}{*}{0.0190}  \\ 
	\cline{2-3}
							   & Minimum leaf size          & 1                 &                          \\ 
	\cline{2-3}
							   & \# of predictors to sample & 8                 &                          \\ 
	\hline
	\multirow{3}{*}{AAoA(rad)} & Number of learners         & 12                & \multirow{3}{*}{0.0797}  \\ 
	\cline{2-3}
							   & Minimum leaf size          & 1                 &                          \\ 
	\cline{2-3}
							   & \# of predictors to sample & 12                &                          \\ 
	\hline
	\multirow{3}{*}{EAoA(rad)} & Number of learners         & 17                & \multirow{3}{*}{0.0220}  \\ 
	\cline{2-3}
							   & Minimum leaf size          & 2                 &                          \\ 
	\cline{2-3}
							   & \# of predictors to sample & 12                &                          \\ 
	\hline
	\multirow{3}{*}{RSS(dB)}   & Number of learners         & 43                & \multirow{3}{*}{0.385}   \\ 
	\cline{2-3}
							   & Minimum leaf size          & 1                 &                          \\ 
	\cline{2-3}
							   & \# of predictors to sample & 9                 &                          \\ 
	\hline
	\multirow{3}{*}{Delay(ns)} & Number of learners         & 255               & \multirow{3}{*}{1.55}    \\ 
	\cline{2-3}
							   & Minimum leaf size          & 6                 &                          \\ 
	\cline{2-3}
							   & \# of predictors to sample & 7                 &                          \\
	\hline
	\end{tabular}
	\end{table}

\subsubsection{Two-step model trained with 50\% of required paths}
We first tune the number of CONV layers in the \#8 1D-CNN and train the newly-built 1D-CNN and six regression tree ensembles with 50\% of required paths. Performances of 1D-CNNs with the number of CONV layers ranging from $5$ to $12$ are presented in Table~\ref{tab:0.5 results}. The performance of the \#12 1D-CNN becomes inferior when it is trained with 50\% of required paths as compared to the \#8 1D-CNN trained with 70\% of required paths. However, when more CONV layers are added to the 1D-CNN, such as \#13-\#15 1D-CNNs, their performances are improved and even surpass that of the \#8 1D-CNN. Thus, the two-step model trained with 50\% of required paths is composed of the \#15 1D-CNN and six regression tree ensembles summarized in Table~\ref{tab:0.5RT}.

\begin{table}[t!]
	\centering
	\caption{Performances of 1D-CNNs Trained with 50\% of Required Paths}
	\label{tab:0.5 results}
	\begin{tabular}{|l|l|l|l|l|l|} 
	\hline
	\#                             & \begin{tabular}[c]{@{}l@{}}\# of\\CONV\\layers \end{tabular} & Recall  & Precision & F-score & \begin{tabular}[c]{@{}l@{}}Cohen's\\kappa \end{tabular}  \\ 
	\hline
	10                             & 5                                                            & 69.79\% & 88.34\%   & 77.98\% & 0.7759                                                                    \\ 
	\hline
	11                             & 7                                                            & 88.41\% & 94.39\%   & 91.30\% & 0.9113                                                                    \\ 
	\hline
	12                             & 9                                                            & 91.52\% & 95.52\%   & 93.47\% & 0.9335                                                                    \\ 
	\hline
	13                             & 10                                                           & 92.64\% & 96.09\%   & 94.33\% & 0.9422                                                                    \\ 
	\hline
	14                             & 11                                                           & 93.24\% & 95.90\%   & 94.55\% & 0.9445                                                                    \\ 
	\hline
	\rowcolor[rgb]{0.8,0.8,0.8} 15 & 12                                                           & 93.35\% & 96.26\%   & 94.79\% & 0.9468                                                                    \\
	\hline
	\end{tabular}
\end{table}

\begin{table}[t!]
	\centering
	\caption{Performances of Regression Tree Ensembles Trained with 50\% of Required Paths.}
	\label{tab:0.5RT}
	\begin{tabular}{|l|l|l|l|} 
	\hline
	Path Parameters            & \multicolumn{2}{l|}{Optimized Hyperparameters} & Val. RMSE                \\ 
	\hline
	\multirow{3}{*}{AAoD(rad)} & Number of learners & 10                        & \multirow{3}{*}{0.0817}  \\ 
	\cline{2-3}
							   & Minimum leaf size  & 4                         &                          \\ 
	\cline{2-3}
							   & \# of predictors   & 11                        &                          \\ 
	\hline
	\multirow{3}{*}{EAoD(rad)} & Number of learners & 10                        & \multirow{3}{*}{0.0238}  \\ 
	\cline{2-3}
							   & Minimum leaf size  & 2                         &                          \\ 
	\cline{2-3}
							   & \# of predictors   & 12                        &                          \\ 
	\hline
	\multirow{3}{*}{AAoA(rad)} & Number of learners & 133                       & \multirow{3}{*}{0.110}   \\ 
	\cline{2-3}
							   & Minimum leaf size  & 1                         &                          \\ 
	\cline{2-3}
							   & \# of predictors   & 12                        &                          \\ 
	\hline
	\multirow{3}{*}{EAoA(rad)} & Number of learners & 48                        & \multirow{3}{*}{0.0278}  \\ 
	\cline{2-3}
							   & Minimum leaf size  & 1                         &                          \\ 
	\cline{2-3}
							   & \# of predictors   & 12                        &                          \\ 
	\hline
	\multirow{3}{*}{RSS(dB)}   & Number of learners & 101                       & \multirow{3}{*}{0.565}   \\ 
	\cline{2-3}
							   & Minimum leaf size  & 1                         &                          \\ 
	\cline{2-3}
							   & \# of predictors   & 12                        &                          \\ 
	\hline
	\multirow{3}{*}{Delay(ns)} & Number of learners & 37                        & \multirow{3}{*}{1.33}    \\ 
	\cline{2-3}
							   & Minimum leaf size  & 3                         &                          \\ 
	\cline{2-3}
							   & \# of predictors   & 6                         &                          \\
	\hline
	\end{tabular}
	\end{table}

As compared to Table~\ref{tab:0.7RT}, performances of six regression tree ensembles trained with 50\% of required paths worsen, with RMSE significantly rised by 55\%, 46\%, 38\%, 26\%, 47\%, -14\% in AAoD(rad), EAoD(rad), AAoA(rad), EAoA(rad), RSS(dB), delay(ns), respectively.

\subsubsection{Two-step model trained with 30\% of required paths}
Similarly, we tune the number of layers of the \#8 1D-CNN and train this two-step model with 30\% of required paths. Table~\ref{tab:0.3 results} demonstrates that the performance of the 1D-CNN with $9$ CONV layers (\#16) is comparable to that with $12$ CONV layers (\#17). Thus, the two-step model trained with 30\% of required paths consists of the \#16 1D-CNN and six regression tree ensembles summarized in Table~\ref{tab:0.3RT}.

\begin{table}[t!]
	\centering
	\caption{Performances of 1D-CNNs Trained with 30\% of Required Paths}
	\label{tab:0.3 results}
	\begin{tabular}{|l|l|l|l|l|l|} 
	\hline
	\#                             & \begin{tabular}[c]{@{}l@{}}\# of\\CONV\\layers\end{tabular} & Recall  & Precision & F-score & \begin{tabular}[c]{@{}l@{}}Cohen's\\kappa\end{tabular}  \\ 
	\hline
	\rowcolor[rgb]{0.8,0.8,0.8} 16 & 9                                                           & 93.62\% & 96.81\%   & 95.19\% & 0.9510                                                  \\ 
	\hline
	17                             & 12                                                          & 93.85\% & 96.38\%   & 95.10\% & 0.9500                                                  \\
	\hline
	\end{tabular}
\end{table}

\begin{table}[t!]
	\centering
	\caption{Performances of Regression Tree Ensembles Trained with 30\% of Required Paths.}
	\label{tab:0.3RT}
	\begin{tabular}{|l|l|l|l|} 
	\hline
	Path Parameters            & \multicolumn{2}{l|}{Optimized Hyperparameters} & Val. RMSE                \\ 
	\hline
	\multirow{3}{*}{AAoD(rad)} & Number of learners & 95                        & \multirow{3}{*}{0.0688}  \\ 
	\cline{2-3}
							   & Minimum leaf size  & 1                         &                          \\ 
	\cline{2-3}
							   & \# of predictors   & 12                        &                          \\ 
	\hline
	\multirow{3}{*}{EAoD(rad)} & Number of learners & 47                        & \multirow{3}{*}{0.0282}  \\ 
	\cline{2-3}
							   & Minimum leaf size  & 3                         &                          \\ 
	\cline{2-3}
							   & \# of predictors   & 12                        &                          \\ 
	\hline
	\multirow{3}{*}{AAoA(rad)} & Number of learners & 11                        & \multirow{3}{*}{0.1561}  \\ 
	\cline{2-3}
							   & Minimum leaf size  & 3                         &                          \\ 
	\cline{2-3}
							   & \# of predictors   & 12                        &                          \\ 
	\hline
	\multirow{3}{*}{EAoA(rad)} & Number of learners & 11                        & \multirow{3}{*}{0.0370}  \\ 
	\cline{2-3}
							   & Minimum leaf size  & 1                         &                          \\ 
	\cline{2-3}
							   & \# of predictors   & 9                         &                          \\ 
	\hline
	\multirow{3}{*}{RSS(dB)}   & Number of learners & 10                        & \multirow{3}{*}{0.6686}  \\ 
	\cline{2-3}
							   & Minimum leaf size  & 1                         &                          \\ 
	\cline{2-3}
							   & \# of predictors   & 9                         &                          \\ 
	\hline
	\multirow{3}{*}{Delay(ns)} & Number of learners & 107                       & \multirow{3}{*}{2.4571}  \\ 
	\cline{2-3}
							   & Minimum leaf size  & 26                        &                          \\ 
	\cline{2-3}
							   & \# of predictors   & 10                        &                          \\
	\hline
	\end{tabular}
	\end{table}

Compared to Table~\ref{tab:0.5RT}, in general, performances of six regression tree ensembles trained with 30\% of required paths in Table~\ref{tab:0.3RT} deteriorate, with RMSE notably rised by $-15\%, 18\%, 42\%, 33\%, 18\%, 85\%$ in AAoD(rad), EAoD(rad), AAoA(rad), EAoA(rad), RSS(dB), delay(ns), respectively. 

\subsubsection{Overall performance of DNN-based path estimation models}
Comparing the performance of the one-step model(0.7:0.3) and the two-step model(0.7:0.3) in Table~\ref{tab:overallperf} indicates only a slight improvement in RMSE, but Fig.~\ref{fig:treeAoD} and Fig.~\ref{fig:FFNNAoD} clearly show the noticeable fitting performance enhancement of the two-step model over the one-step model.
Besides, though the two-step model trained with 30\% of required paths learns less from the UT-level CSI given by ray tracing, it achieves comparable and even better performance in path parameter prediction comparing to the two-steo model trained with 50\% and 70\% of required paths, as shown in Table~\ref{tab:overallperf}.

\begin{table*}[t!]
	\centering
	\caption{Overall Performances of DNN-based Path Estimation Models.}
	\label{tab:overallperf}
	\begin{tabular}{|l|c|c|c|c|c|c|}
	\hline
	\diagbox{Model}{RMSE}{Param.} & AAoD(rad) & EAoD(rad) & AAoA(rad) & EAoA(rad) & RSS(dB) & Delay(ns) \\ \hline
	0.3:0.7(two-step)                                  & 0.0858    & 0.0216    & 0.0799    & 0.0216    & 2.4069  & 2.218     \\ \hline
	0.5:0.5(two-step)                                  & 0.0981    & 0.0217    & 0.0839    & 0.0215    & 2.4693  & 2.267     \\ \hline
	0.7:0.3(two-step)                                  & 0.0935    & 0.0212    & 0.0871    & 0.0213    & 2.6261  & 2.256     \\ \hline
	0.7:0.3(one-step)                                  & 0.1062    & 0.0141    & 0.1323    & 0.0142    & 3.2771  & 3.116     \\ \hline
	\end{tabular}
\end{table*}

\subsubsection{Comparison between ray tracing and DNN-based path estimation model}
\begin{figure}[t!]
	\centering
	\includegraphics[width=0.6\linewidth]{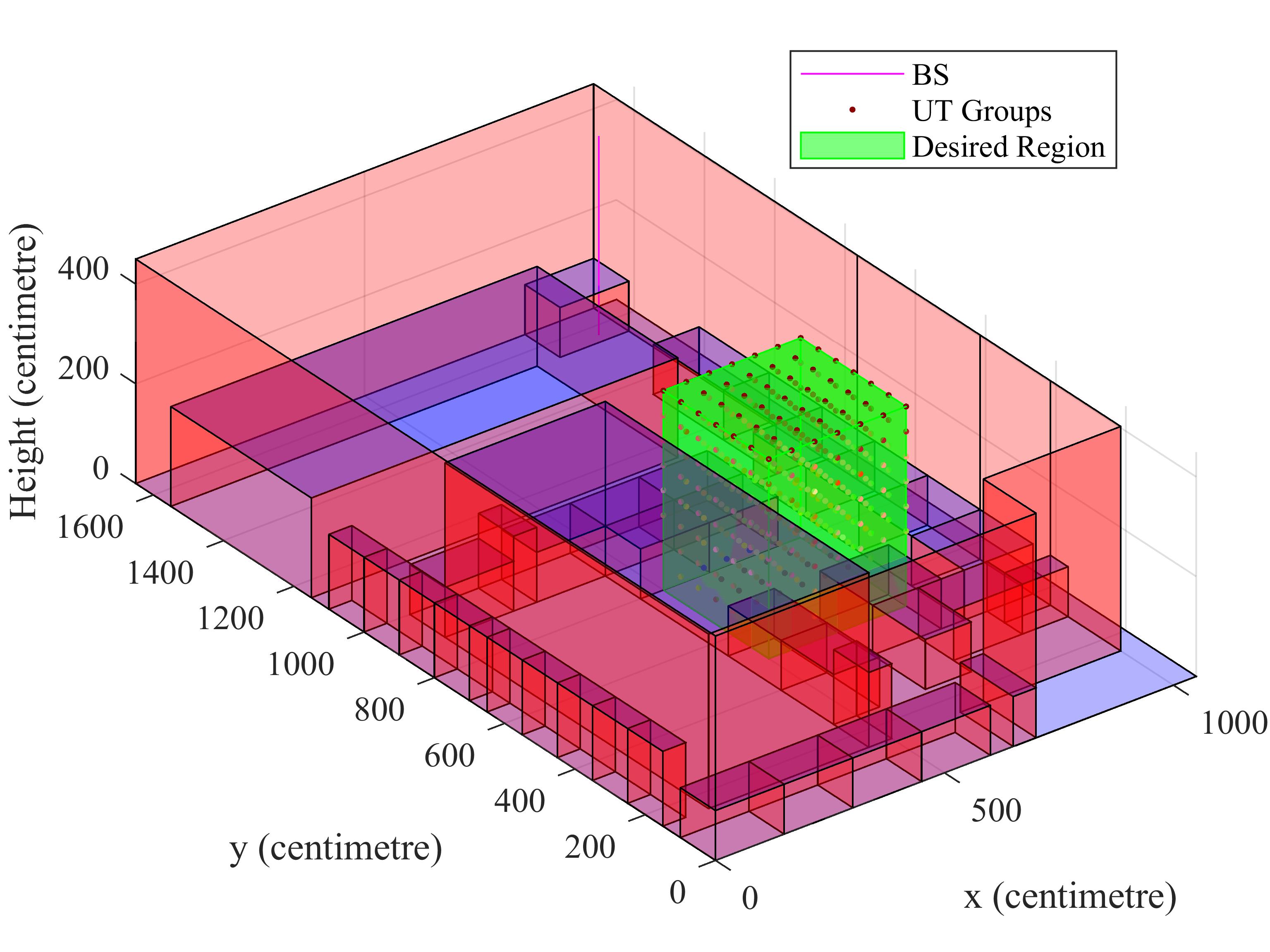}
	\caption{UT locations with unknown UT-level CSI.}
	\label{fig:MLregion1}
\end{figure}

\begin{figure}[t!]
	\centering
	\subfloat[Ray tracing.]{
		\includegraphics[width=0.6\linewidth]{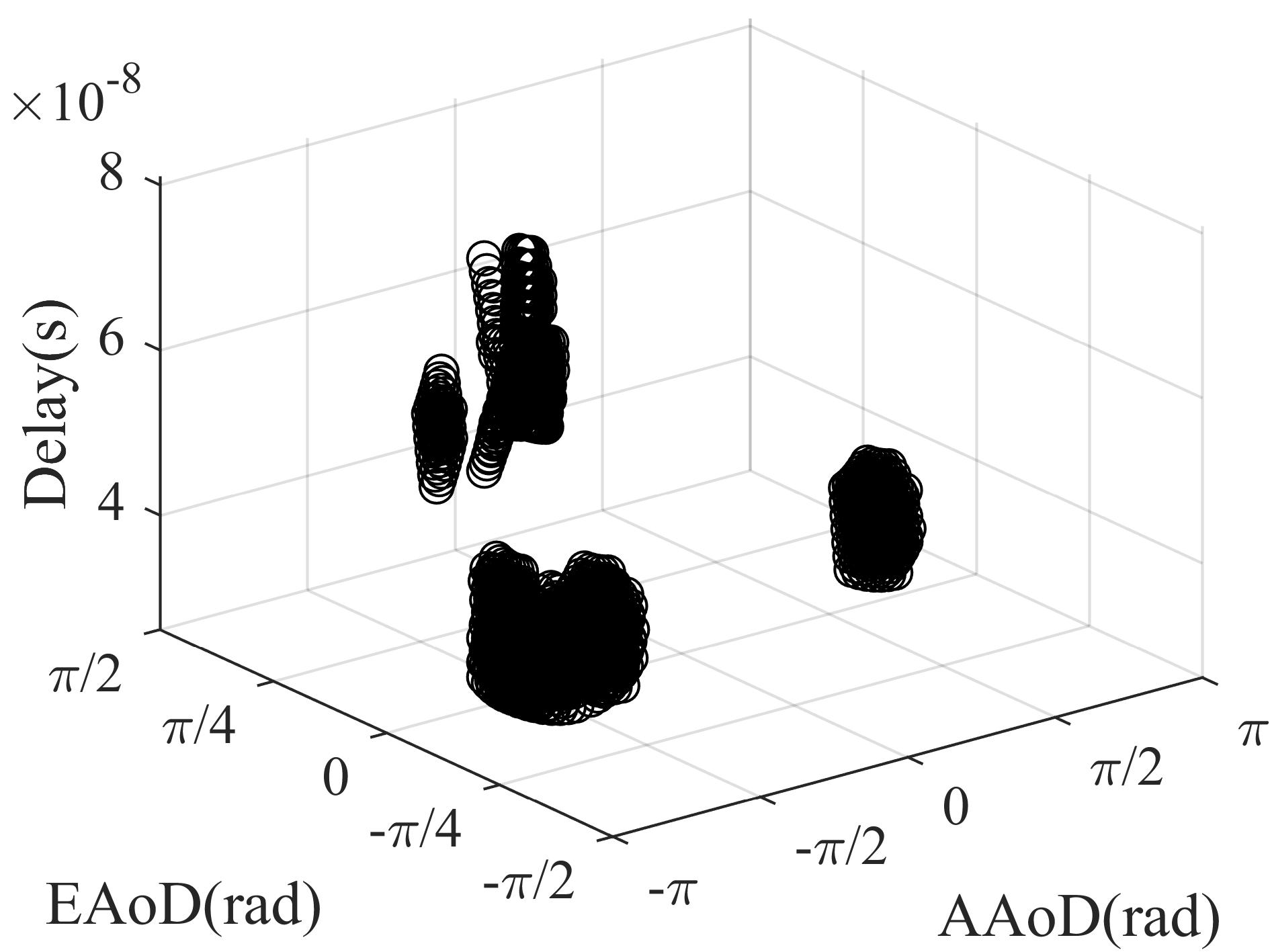}
		\label{fig:MPCRT}
	}
	\hfill
	\subfloat[Two-step model (0.7:0.3).]{
		\includegraphics[width=0.6\linewidth]{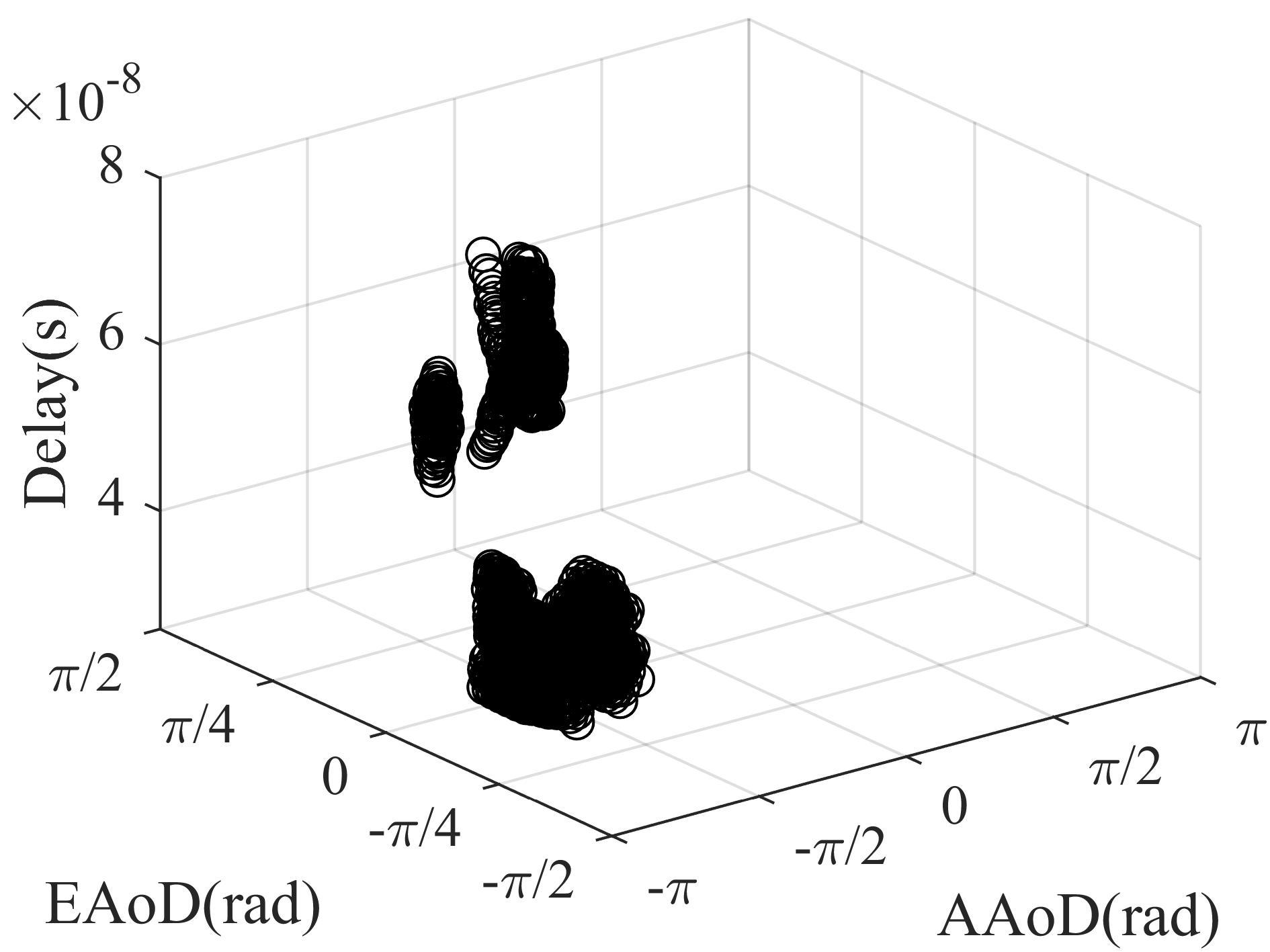}
		\label{fig:MPCML1}
	}
	\hfill
	\subfloat[Two-step model (0.5:0.5).]{
		\includegraphics[width=0.6\linewidth]{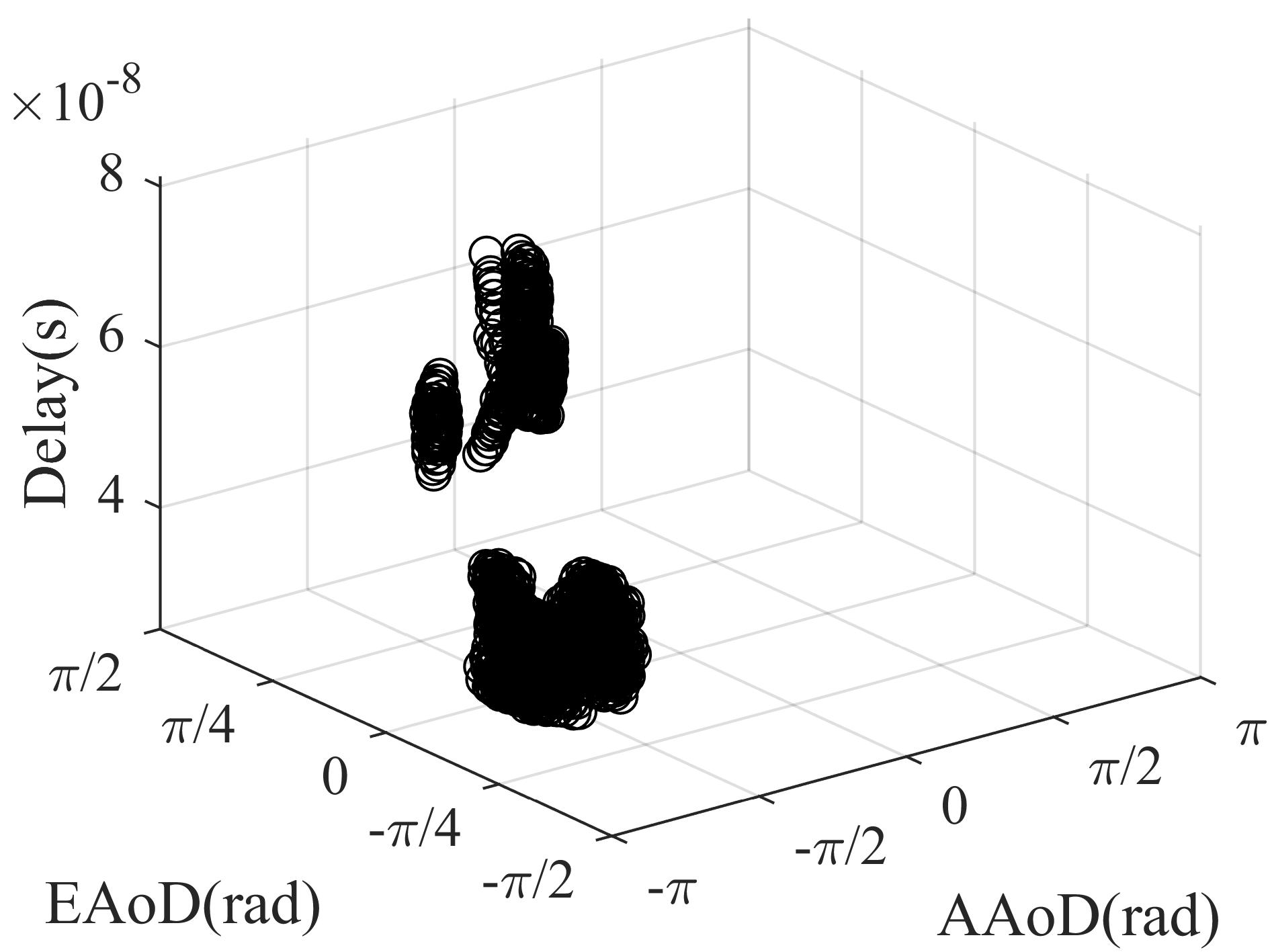}
		\label{fig:MPCML2}
	}
	\hfill
	\subfloat[Two-step model (0.3:0.7).]{
		\includegraphics[width=0.6\linewidth]{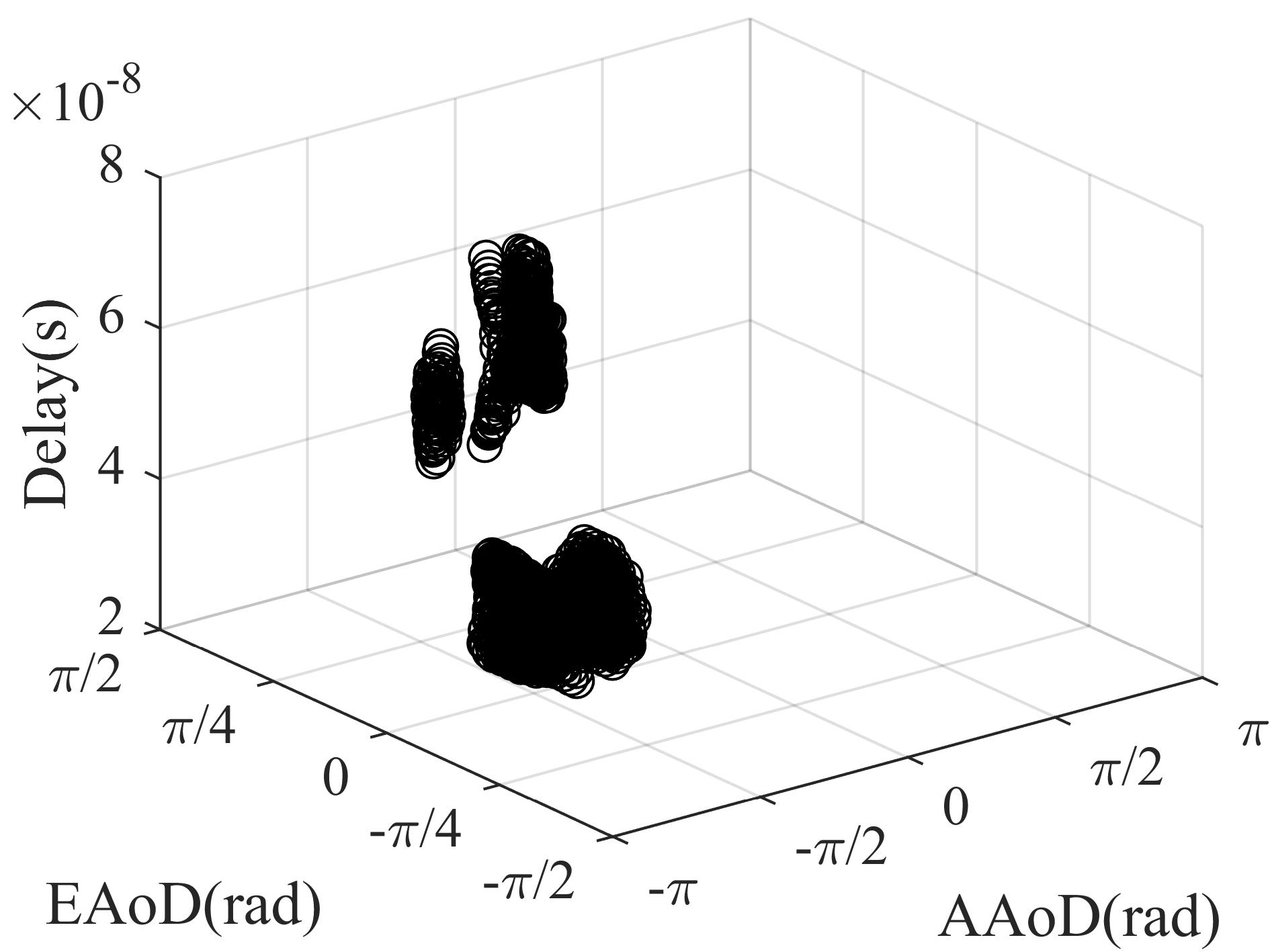}
		\label{fig:MPCML3}
	}
	\hfill
	\caption{Paths for the selected service area obtained by ray tracing and DNN-based path estimation models. }
	\label{fig:MPCMLRT}
\end{figure}

There are $73,336$ paths for $356$ feasible UT locations required for the channel estimation within the desired space in Fig.~\ref{fig:MLregion1}. The two-step model (0.7:0.3) generates $1,325$ existent paths and $72,011$ non-existent paths. The two-step model (0.5:0.5) generates $1,350$ existent paths and $71,986$ non-existent paths. The two-step model (0.3:0.7) generates $1354$ existent paths and $71,982$ non-existent paths. As a reference, the ray tracing algorithm produces $1,715$ existent paths and $71,621$ non-existent paths. The AoDs of existent paths obtained by three two-step models and by ray tracing are presented in Fig.~\ref{fig:MPCMLRT}. Obviously, most of existent paths for unknown feasible UT locations can be accurately predicted by the two-step model except some paths having positive AAoD.

The relative times on average to acquire paths for one UT location by ray tracing and three two-step models are shown in Table~\ref{tab:MLsimulation}. Compared with the time taken by ray tracing, the time taken by three two-step models is reduced by $26.4\% $(0.7:0.3), $46\%$ (0.5:0.5) and $65.8\%$ (0.3:0.7), respectively. Therefore, by means of deep neural networks, the UT-level CSI required for the channel estimation is produced in a much shorter time with acceptable errors.
\begin{table}[t!]
	\centering
	\caption{Relative Time of Path Esimation for One UT Location by Ray Tracing, Two-step Models with 70\%, 50\% and 30\% Training Paths.}
	\label{tab:MLsimulation}
	\begin{tabular}{|l|l|l|}
		\hline
		Algorithm      & \# of existent/ non-existent paths & Relative time \\ \hline
		Ray tracing    & 1715/71621                         & 1             \\ \hline
		Model(0.7:0.3) & 1325/72011                         & 0.736         \\ \hline
		Model(0.5:0.5) & 1350/71986                         & 0.540         \\ \hline
		Model(0.3:0.7) & 1354/71982                         & 0.342         \\ \hline
		\end{tabular}
\end{table}

\section{Sum-Rate Performance}
Throughout this section, the proposed channel estimation model is utilized for the MU-mMIMO system illustrated in Fig.~\ref{fig:systemModel}.
	The sum-rate performance of two-stage AB-HP and single-stage FDP techniques is investigated considering the indoor scenario specified in Table \ref{tab:SS}.
Also, Table~\ref{table:IllustrativeResults} summarizes the simulation setup, where the BS is equipped with a uniform rectangular array (URA) having  ${M=16\times16=256}$ antennas.

\begin{table}[!t]
	\caption{Sum-rate Performance Simulation Parameters.}
	\label{table:IllustrativeResults}
	\centering
	\begin{tabular}{|c|c|}
		\hline
		\# of antennas at BS \cite{Report_5G_UMi_UMa_Rel16} & $M=16\times 16$      \\ \hline
		%		{Carrier frequency} \cite{Report_5G_Macro_PL}& $2$ GHz      \\ \hline
		{Channel bandwidth} \cite{Report_5G_Macro_PL}& $10$ kHz      \\ \hline
		{Noise PSD} \cite{Report_5G_Macro_PL}& $-174$ dBm/Hz      \\ \hline
		%		{Transmit power per RRH} \cite{Report_5G_Macro_PL}& $P_T=46$ dBm      \\ \hline
		{Path loss exponent}& $\eta=2.1$      \\ \hline
		{\# of UTs deployed in each group} & $K_g=\frac{K_a}{G_a}$      \\ \hline
		{\# of paths} & $P=10$      \\ \hline
		Antenna spacing (in wavelength) & ${0.5}$\\ \hline
		{\# of network realizations} & $2000$      \\ \hline
	\end{tabular}
\end{table}

\begin{figure}[t!]
	\centering
	\includegraphics[width=\columnwidth]{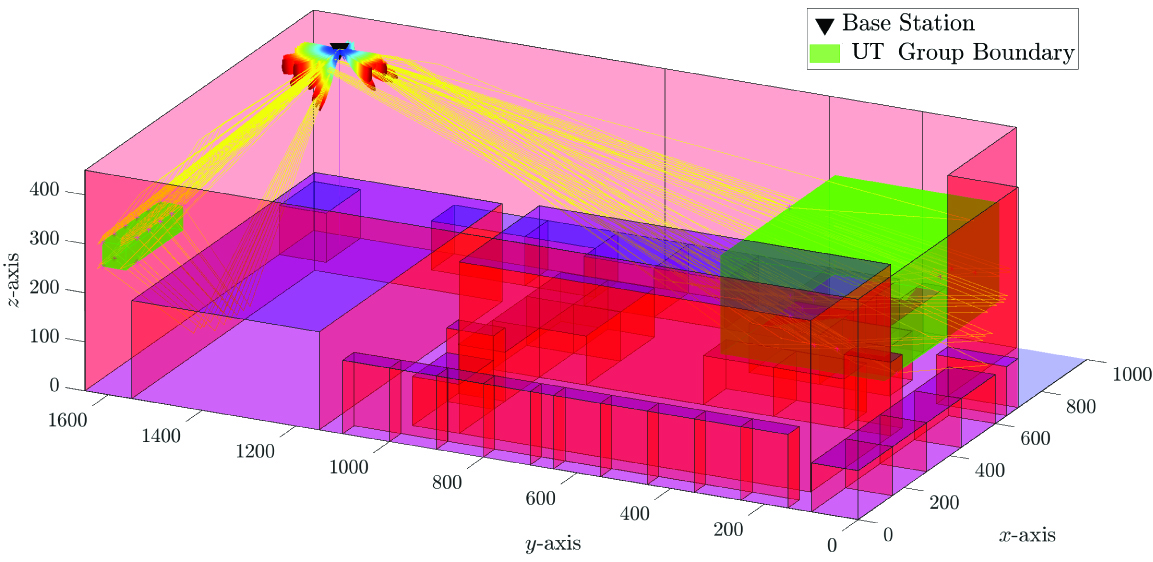}
	\caption{Indoor scenario for $G_a=2$ UT groups with the corresponding beams obtained by the RF beamformer.}
	\label{fig:indoor}
	\vspace{-2ex}
\end{figure}

In Fig.~\ref{fig:indoor}, there are $G_a=2$ UT zones in the service area, where the one on the left has four clusters and the one on the right has five clusters.
	Here, the RF beamformer of the AB-HP technique generates the corresponding beams by using the UT-group CSI of each UT zone obtained via the ray tracing.
It is shown that the generated beams are well-aligned with the multiple paths and UT zones.
	
	AB-HP first builds the RF beamformer ${\bf F}\in\mathbb{C}^{256\times 22}$, which needs $N_{RF}=22$ RF chains.
Then, the BB precoder is generated via the reduced-size effective channel matrix $\bm{\mathcal{H}}={\bf H }\bf{F}\in\mathbb{C}^{K_a\times 22}$ seen from the BB-stage.
On the other hand, the conventional single-stage FDP requires ${N_{RF}=M=256}$ RF chains and the full channel matrix ${\bf H}\in \mathbb{C}^{K_a\times 256}$.
Thus, the AB-HP technique utilizing the proposed channel modeling decreases the CSI overhead size by $91.4\%$ compared to FDP and the number of RF chains are remarkably reduced from $256$ to $22$.

\begin{figure}[t!]
	\centering
	\includegraphics[width=\columnwidth]{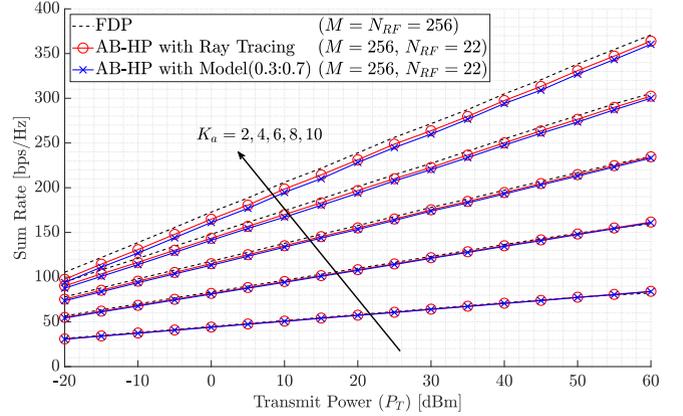}
	\caption{Sum-rate comparison of FDP and AB-HP techniques, where $K_a=2,4,6,8,10$.}
	\label{fig:sumrate}
\end{figure}

Fig.~\ref{fig:sumrate} demonstrates the sum-rate performance of the FDP and AB-HP techniques obtained via (\ref{eq:SINR}) and (\ref{eq:Rate}). Here, we assume that $K_a=2,4,6,8,10$ UTs are equally distributed to $G_a=2$ groups (i.e., $K_g=\frac{K_a}{G_a}$). 
	In order to compare the accuracy of ray tracing and two-step model (0.3:0.7) on the sum-rate performance, the RF beamformer of AB-HP is also fed by the two-step model (0.3:0.7) in addition to ray tracing.
Results show that AB-HP with ray tracing can provide a comparable sum-rate performance with respect to the conventional FDP. For instance, at $P_T=40$~dBm, the performance degradation is only $0.4$ bps/Hz ($7.2$ bps/Hz) for $K_a=2$ ($K_a=10$) users.
	Moreover, AB-HP with ray tracing achieves $99.1\%$ ($97.8\%$) of sum-rate obtained by FDP for $K_a=2$ ($K_a=10$) users, while, AB-HP greatly relaxes the CSI overhead and the hardware cost/complexity as mentioned above.
Furthermore, AB-HP with the two-step model (0.3:0.7) also closely converge the sum-rate performance of AB-HP with ray tracing, whereas, the two-step model decreases the time for the path prediction by $65.8\%$ (please see Table \ref{tab:MLsimulation}).
	For example, two-step model (0.3:0.7) provides $99.6\%$ ($99.2\%$) of the sum-rate performance achieved by ray tracing model for $K_a=2$ ($K_a=10$) users.

\section{Conclusion}
A deep learning based channel estimation approach using 3D geospatial data, the 1D-CNN and the proposed FCM for MU-mMIMO hybrid precoding RF beamformer has been developed. It provides the RF beamformer with offline UT-group CSI for different UT zones in the service area, which significantly reduces the online CSI overhead by 91.4\%. Using a DNN-based technique and ray tracing for UT-level CSI acquisition further shortens the measurement time by 65.8\%, compared to only ray tracing. Meanwhile, the UT-group CSI of each UT zone is robust to the imperfect UT-level CSI produced by the DNN-based technique because of the fuzziness introduced into the clusters by the proposed FCM. 
Finally, considering the MU-mMIMO systems, the illustrative results verify that the AB-HP configured with offline UT-group CSI generated by the proposed DNN-based channel estimation technique can successfully achieve a comparable sum-rate performance with the same AB-HP technique with offline UT-group CSI produced by computationally expensive ray tracing and the FDP technique with full-size online CSI.

%\section*{Acknowledgment}
%This article was presented in part in IEEE International Conference on Communications
%(ICC 2021), Montreal, Canada, June 2021 in \cite{ASIL_XIAOYI_ICC}.
\ifCLASSOPTIONcaptionsoff
\newpage
\fi

\bibliography{references}
\bibliographystyle{IEEEtran}

\begin{IEEEbiography}[{\includegraphics[width=1in,height=1.25in,clip,keepaspectratio]{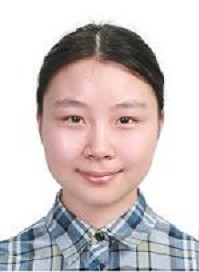}}]{Xiaoyi Zhu} (Student Member, IEEE) 
	received the B.Eng. degree in communication engineering from Wuhan University of Technology, Wuhan, China, in 2018
	and the M.Sc. degree in electrical engineering at McGill University, Montreal, Canada in 2020. 
	Her research interests include wireless channel modelling, 5G millimeter wave propagation, artificial intelligence, massive MIMO, specifically designing AI-based channel estimation methods for massive MIMO hybrid precoding systems.	
\end{IEEEbiography}
\vspace{-63ex}

\begin{IEEEbiography}[{\includegraphics[width=1in,height=1.25in,clip,keepaspectratio]{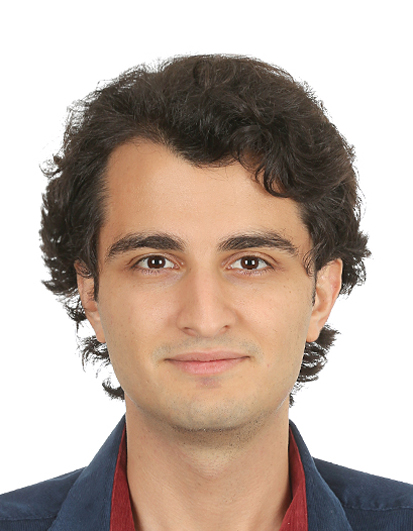}}]{Asil Koc} (Student Member, IEEE) received the B.Sc. degree (Hons.) in electronics and communication engineering and the M.Sc. degree (Hons.) in telecommunication engineering from Istanbul Technical University, Istanbul, Turkey, in 2015 and 2017, respectively. He is currently pursuing the Ph.D. degree in electrical engineering with McGill University, Montreal, QC, Canada. 
	
From 2015 to 2017, he was a Research and Teaching Assistant with the Electronics and Communication Engineering Department, Istanbul Technical University. Since 2017, he has been a Teaching Assistant with the Electrical and Computer Engineering Department, McGill University. His research interests include, but not limited to wireless communications, massive MIMO, full-duplex, spatial modulation, energy harvesting, and cooperative networks. He was a recipient of the Erasmus Scholarship Award funded by the European Union, the McGill Engineering Doctoral Award, the STARaCom Collaborative Grant funded by the FRQNT, and the Graduate Research Enhancement and Travel Award funded by McGill University. 
\end{IEEEbiography}

\newpage

\begin{IEEEbiography}[{\includegraphics[width=1in,height=1.25in,clip,keepaspectratio]{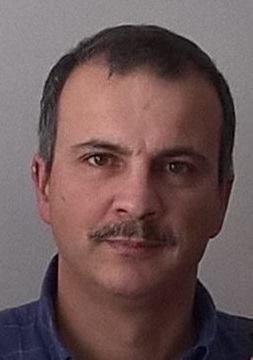}}]{Robert Morawski}
	received the B.Sc. and M.Sc. degrees in electrical and computer engineering from Concordia University, Montreal, QC, Canada, in 1997 and 2000, respectively. He is currently a Research Engineer and a Lab Manager with the Broadband Communications Research Lab, ECE Department, McGill University. His current research interests include the design and implementation of prototype architectures for next-generation wireless communications. 
\end{IEEEbiography}

\vspace{-70ex}

\begin{IEEEbiography}[{\includegraphics[width=1in,height=1.25in,clip,keepaspectratio]{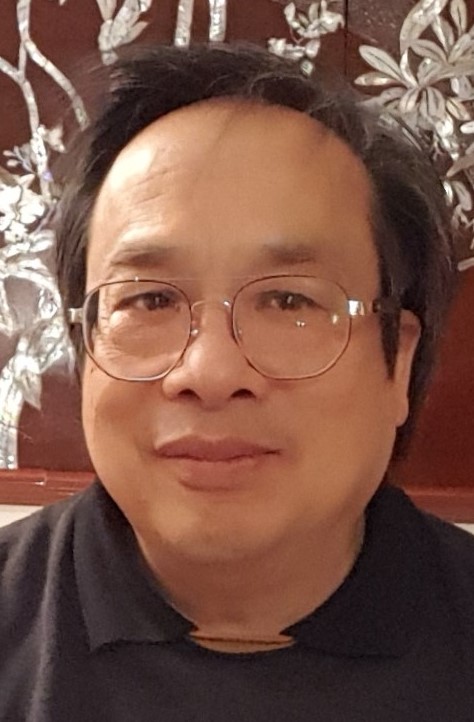}}]{Tho Le-Ngoc}(Life Fellow, IEEE) received the B.Eng. degree in electrical engineering in 1976, the M.Eng. degree in microprocessor applications from McGill University, Montreal, in 1978, and the Ph.D. degree in digital communications from the University of Ottawa, Canada, in 1983.
	
From 1977 to 1982, he was with Spar Aerospace Ltd., Sainte-Anne-de-Bellevue, QC, Canada, involved in the development and design of satellite communications systems. From 1982 to 1985, he was with SRTelecom, Inc., Saint Laurent, QC, Canada, where he developed the new point-to-multipoint DA-TDMA/TDM Subscriber Radio System SR500. From 1985 to 2000, he was a Professor with the Department of Electrical and Computer Engineering, Concordia University, Montreal. Since 2000, he has been with the Department of Electrical and Computer Engineering, McGill University. His research interest includes broadband digital communications. He was a recipient of the 2004 Canadian Award in Telecommunications Research and the IEEE Canada Fessenden Award in 2005. He is a Distinguished James McGill Professor, and a Fellow of the Engineering Institute of Canada, the Canadian Academy of Engineering, and the Royal Society of Canada.
\end{IEEEbiography}

\EOD

\end{document}